 \documentclass[final,5p,times,twocolumn,authoryear]{elsarticle}

\usepackage{amssymb}
\usepackage{lipsum}

\usepackage{subfigure}
\usepackage{subcaption}
\usepackage{verbatim}
\usepackage{multirow}
\usepackage{amsmath}
\usepackage{acronym}
\usepackage{diagbox}
\usepackage{arydshln}
\usepackage{siunitx}    
\usepackage{hyperref}
\usepackage{tabu}
\usepackage{placeins}
\usepackage{xcolor}
\usepackage{ulem}

\journal{Experimental Thermal and Fluid Science}

\acrodef{2D}[2D]{two-dimensional}
\acrodef{3D}[3D]{three-dimensional}
\acrodef{2C}[2C]{two-component}
\acrodef{3C}[3C]{three-component}
\acrodef{2D-3C}[2D-3C]{two-dimensional three-component}
\acrodef{AER}[AER]{Address-Event Representation}
\acrodef{BL}[BL]{boundary layer}
\acrodef{CD}[CD]{Contrast Detection}
\acrodef{CFD}[CFD]{computational fluid dynamics}
\acrodef{DMD}[DMD]{Dynamic Mode Decomposition}
\acrodef{DNS}[DNS]{Direct Numerical Simulation}
\acrodef{DVS}[DVS]{dynamic vision sensing}
\acrodef{DSR}[DSR]{Dynamic Spatial Range}
\acrodef{DVR}[DVR]{dynamic velocity range}
\acrodef{EBIV}[EBIV]{Event-Based Imaging Velocimetry}
\acrodef{EBV}[EBV]{Event-based Vision}
\acrodef{EPOD}[EPOD]
{Extended Proper Orthogonal Decomposition}
\acrodef{FOV}[FOV]{Field of View}
\acrodef{FWHM}[FWHM]{full-width-half-mean}
\acrodef{HFSB}[HFSB]{helium filled soap bubbles}
\acrodef{HS}[HS]{high-speed}
\acrodef{HWA}[HWA]{hot-wire anemometry}
\acrodef{LDA}[LDA]{Laser Doppler Anemometry}
\acrodef{LES}[LES]{large-eddy simulation}
\acrodef{LSE}[LSE]{Linear Stochastic Estimation}
\acrodef{LOR}[LOR]{Low Order Reconstruction}
\acrodef{LPT}[LPT]{Lagrangian Particle Tracking}
\acrodef{PDF}[PDF]{Probability Density function}
\acrodef{PIV}[PIV]{Particle Image Velocimetry}
\acrodef{POD}[POD]
{Proper Orthogonal Decomposition}
\acrodef{PSD}[PSD]{Power Spectral Density}
\acrodef{ppp}[ppp]{particles per pixel}
\acrodef{PTV}[PTV]{Particle Tracking Velocimetry}
\acrodef{PWM}[PWM]{pulse width modulation}
\acrodef{ROM}[ROM]{reduced-order modeling}
\acrodef{RMS}[rms]{root mean square}
\acrodef{RMSE}[RMSE]{root mean square error}
\acrodef{2D-2C-PIV}[2D-2C-PIV]{two-dimensional (planar), two component particle image velocimetry}
\acrodef{ROI}[ROI]{region of interest}
\acrodef{SVD}[SVD]{singular value decomposition}
\acrodef{TBL}[TBL]{turbulent boundary layer}
\acrodef{TKE}[TKE]{turbulent kinetic energy}

\begin{document}

\begin{frontmatter}



\title{An efficient offline sensor placement method for flow estimation}


\author[a]{Junwei Chen}\corref{mycorrespondingauthor}
\cortext[mycorrespondingauthor]{Corresponding author}
\ead{junwei.chen@uc3m.es}
\author[a]{Marco Raiola}
\ead{mraiola@ing.uc3m.es}
\author[a]{Stefano Discetti}
\ead{sdiscett@ing.uc3m.es}
\address[a]{Department of Aerospace Engineering, Universidad Carlos III de Madrid, Avda. Universidad 30, 28911, Leganes, Spain}


\begin{abstract}
We present an efficient method to optimize sensor placement for flow estimation using sensors with time-delay embedding in advection-dominated flows. Our solution allows identifying promising candidates for sensor positions using solely preliminary flow field measurements with non-time-resolved \ac{PIV}, without introducing physical probes in the flow. Data-driven estimation in advection-dominated flows often exploits time-delay embedding to enrich the sensor information for the reconstruction, i.e. it uses the information embedded in probe time series to provide a more accurate estimation. Optimizing the probe position is the key to improving the accuracy of such estimation. Unfortunately, the cost of performing an online combinatorial search to identify the optimal sensor placement in experiments is often prohibitive. We leverage the principle that, in advection-dominated flows, rows of vectors from \ac{PIV} fields embed similar information to that of probe time series located at the downstream end of the domain. We propose thus to optimize the sensor placement using the row data from non-time-resolved \ac{PIV} measurements as a surrogate of the data a real probe would actually capture in time. This optimization is run offline and requires only one preliminary experiment with standard \ac{PIV}. Once the optimal positions are identified, the probes can be installed and operated simultaneously with the \ac{PIV} to perform the time-resolved field estimation. We show that the proposed method outperforms equidistant positioning or greedy optimization techniques available in the literature.
\end{abstract}



\begin{keyword}
Particle Image Velocimetry \sep Proper Orthogonal Decomposition \sep Data-driven estimation



\end{keyword}

\end{frontmatter}




\section{Introduction}
\label{sec:introduction}
\acf{PIV} is referred to as ``time-resolved'' when the repetition rate of illumination and imaging system is fast enough to enable measurements of the flow dynamics with sufficient temporal resolution. Time-resolved \ac{PIV} has now become a widespread technique in fluid mechanics \citep{beresh2021time}, unlocking a detailed view of flow dynamics and the possibility of extracting acceleration and pressure fields \citep{van2013piv}. Time-resolved flow fields allow the description of the dynamics of coherent structures, thus they are a key enabler for low-order modelling and control. However, technological limitations prevent access to time-resolved flow descriptions if the flow is sufficiently fast. 

As a cost-effective and flexible alternative, high-repetition-rate point probes (such as hot wires and microphones) can be combined with simultaneous low-repetition-rate \ac{PIV} measurements. Data-driven techniques can then be leveraged to estimate flow fields at a temporal resolution up to that of point probe data. The reconstruction problem is ill-posed: few sensors are used to estimate flow fields composed of a much larger number of vectors. However, it can still be resolved with reasonable accuracy according to the following two observations. First, often the flow dynamics evolve on low-dimensional attractors. Dimensionality-reduction techniques, such as \acl{POD} \citep[\acs{POD},][]{berkooz1993proper}, manifold learning \citep{lee2007nonlinear,meilua2024manifold}, or more recently autoencoders \citep{kramer1991nonlinear,bank2023autoencoders}, can be leveraged to identify low-dimensional intrinsic variables for the velocity fields. Second, assumptions on flow physics (for instance dominance of advection) may allow time-delay embedding, thus further increasing the available information. Along this line, approaches based on multi-time delay with \acl{LSE} \citep[\acs{LSE},][]{ewing1999examination, tinney2006spectral, tinney2008low} or Extended \acs{POD} \citep[\acs{EPOD},][]{boree2003extended,sicot2012wall, tu2013integration, hosseini2015sensor, discetti2018estimation, discetti2019characterization} have been successfully implemented.
Some of the earlier approaches have been formulated in Ref. \cite{clark2014general} in the framework of linear least-square estimation. The work provided interesting guidelines and practical recommendation for the use of stochastic estimation from sensors. The aforementioned approaches leverage the correlation between data collected by the sensors and field data. Consequently, the reconstructed fields are filtering out information that is not correlated with the data measured by the sensors.
Additionally, the estimated time-resolved flow fields can be exploited to retrieve additional field information. Recently, it was demonstrated that pressure fields can also be obtained from the estimated velocity fields by leveraging compliance with the Navier-Stokes equations and setting proper boundary conditions \citep{chen2022pressure,morenosoto2024complete}.

The sensor positioning has a crucial impact on the quality of the flow estimation. Proper placement improves the accuracy of flow field predictions and/or allows reducing the number of required sensors, thus simplifying the measurement setup. The main difficulty is that sensor placement requires solving a NP-hard (non-deterministic polynomial-time hard) combinatorial problem. Several methods have been proposed to solve it with an affordable computational cost. Remarkable examples include: QR-pivoting \citep{manohar2018data}, building upon the discrete empirical interpolation method \citep{drmac2016new,chaturantabut2010nonlinear}, Shannon's entropy \citep{papadopoulou2014hierarchical}, variance inflation factor \citep{nickels2020low}, characteristics of Fisher Information matrix \citep{nakai2021effect} and reinforcement learning \citep{paris2021robust}. The main issue is that these methods do not account for the intrusiveness of the sensor. This is a key aspect, which most often leads to sensor placement at walls, or at the downstream end of the domain if advection is dominant. 

In this work, we treat optimal sensor placement in this latter case, in which there is a main advection direction and point probes can be placed right at the downstream end of the domain. This introduces further complexity to the sensor optimization problem. The reconstruction accuracy, in this scenario,  depends indeed also on the effectiveness of time-delay embedding in the selected locations. Such analysis can only be done a posteriori (i.e. after probe positioning and data acquisition). This would require performing the experiment multiple times with different probe positioning and searching for the best location. Even with a small handful of probes, this implementation is unaffordable. 

In this paper, we propose an offline solution for sensor positioning, based solely on non-time-resolved \ac{PIV} fields, i.e. without placing physical probes in the first preparatory run. Synthetic point probes are extracted from \ac{PIV} fields to simulate the effect of including time history when placing the real ones. Using data from the preparatory \ac{PIV} run, the flow field is reconstructed with different combinations of sensor locations on the downstream edge of the domain to identify the optimal positioning with the highest reconstruction accuracy. Once the most performing sensor placement is identified, the estimation experiment can be carried out by placing the physical probes and measuring the velocity fields simultaneously, and time-resolved flow field can be reconstructed from the measurement using \ac{PIV} and sensors together.

\acused{POD}
\acused{EPOD}

The paper is structured as follows. First, the methodology is described in Sec. \ref{sec:methods}, including the estimation of \ac{EPOD}, the process of online and offline sensor placement optimization, sensor formations, as well as block-pivoted QR for sensor placement. Then the validation test cases are illustrated (Sec. \ref{sec:validation}) and the results are discussed (Sec. \ref{sec:results}). Finally, the conclusions are drawn.

\section{Methods}
\label{sec:methods}

\subsection{Flow field estimator based on Extended Proper Orthogonal Decomposition}
\label{sec:EPOD}

For the flow field estimation from probes, we use \ac{EPOD}, a fast and reliable method that allows efficient exploration of sensor placements. While this paper focuses on \ac{EPOD}, the same approach can extend to other estimators, e.g. those based on recurrent neural networks \citep{deng2019time}.

The \ac{EPOD} \citep{boree2003extended} is a linear tool to establish the correlation between different quantities. In this framework, it is used to project the velocity field data from snapshot \ac{PIV} on \ac{POD} modes of the high-repetition-rate probe data at the same time instants. This allows the establishment of correlations between temporal modes of velocity field and probe data. The correlation can be leveraged then to estimate time coefficients of the velocity fields solely from probe data, thus delivering flow estimation.

This method is computationally affordable and provides acceptable estimation in simple flows (i.e. with a sufficiently compact \ac{POD} spectrum and good evidence of correlation between sensor and field data, such as in shedding-dominated flows). The field estimation methodology is briefly described in this section. The reader can refer to \cite{discetti2018estimation} for a more detailed formulation. Furthermore, neural networks could also be used for this task \citep{erichson2020shallow}, although at the expense of a higher computational cost.

Consider a dataset of $n_t$ snapshots of the fluctuating velocity field. The velocity is measured on $n_p$ grid points. A snapshot matrix $\mathbf{U}$ is built by arranging each snapshot as a row. Following the snapshot method from \cite{sirovich1987turbulence}, the \ac{POD} is obtained by decomposing $U$ with an economy-size \ac{SVD}, i.e.
\begin{equation}
    \mathbf{U}=\mathbf{\Psi}\mathbf{\Sigma}\mathbf{\Phi}^T
    \label{eqn1}
\end{equation}
It is assumed for simplicity $n_t<n_c n_p$ (being $n_c$ the number of velocity components measured at each grid point). The decomposition of Eq. \ref{eqn1} generates two square unitary matrices: $\mathbf{\Psi}$, of size $n_t\times n_t$, containing the temporal modes; $\mathbf{\Phi}^T$ containing the spatial modes, of size $n_cn_p \times n_cn_p$ (being it actually of size $n_t \times n_cn_p$ in the economy-size \ac{SVD} for the case of $n_t<n_c n_p$). The diagonal matrix $\mathbf{\Sigma}$ contains the singular values $\sigma_i$ sorted by their magnitude.

We apply the same decomposition to datasets acquired at the same time instants of velocity fields and probe measurements. As mentioned above, the number of probes is artificially increased by a multi-time-delay embedding. This approach is also referred to as Multichannel Singular Spectrum Analysis \citep{ghil2002advanced}. In particular, being the probes generally located at the downstream end of the domain, for each physical probe and at each time instant we assign a portion of the time-resolved sequence of the probe of $q$ samples corresponding to time instants past the one of each corresponding \ac{PIV} snapshot. This multi-time-delay embedding results in a probe snapshot matrix $\mathbf{U}_{pr}$ with $n_t$ rows and $n_{tt}=(s\times n_{c_{pr}})\times q$ columns (being $s$ the number of probes and $n_{c_{pr}}$ the number of flow quantity components measured by the probes). This matrix undergoes the same decomposition of the velocity snapshot matrix, resulting in:
\begin{equation}
    \mathbf{U}_{pr} = \mathbf{\Psi}_{pr}\mathbf{\Sigma}_{pr}\mathbf{\Phi}_{pr}^T
    \label{eqn2}
\end{equation}

The correlation matrix of temporal modes is thus built as $\mathbf{\Xi}=\mathbf{\Psi}_{pr}^T\mathbf{\Psi}$. The matrix $\mathbf{\Xi}$ establishes the correlation between the velocity field and probe temporal modes.

The estimation of the velocity field at a generic time instant $t^*$ is based on three steps: (1) computation by projection of the temporal modes of the probes corresponding to $t^*$; (2) estimation of the temporal modes of the velocity fields $\mathbf{\psi}_{est}$ leveraging their relation with the temporal modes of the probes through the correlation matrix $\mathbf{\Xi}$; (3) calculation of the velocity field using the estimated temporal modes and the spatial modes from the original decomposition. It can be shown that this can be condensed in the following relation:
\begin{equation}
    \mathbf{u}_{est}=\mathbf{u}_{se}\mathbf{\Phi}_{pr}\mathbf{\Sigma}_{pr}^{-1}\mathbf{\Xi}\mathbf{\Sigma}\mathbf{\Phi}^T=\mathbf{\psi}_{est}\mathbf{\Sigma}\mathbf{\Phi}^T
    \label{eqn4}
\end{equation}
with $\mathbf{u}_{se}$ being a vector containing delay-embedded probe data arranged with the same procedure used to build the probe snapshot matrix. Further improvement of the estimation can be obtained by filtering out spurious contributions due to weakly-correlated features, following the method proposed by \cite{discetti2018estimation}. 

In this paper, the \ac{EPOD} will be used for both training and testing. During the training phase, it will establish the relationship between the velocity field and sensor data using non-time-resolved data. In the testing phase, the estimated time-resolved field will be compared to the reference field.

\subsection{Online vs. offline sensor placement optimization}
\label{subsec:offline_sensor_placement}
\begin{figure*}
\centering
\includegraphics[scale=0.38]{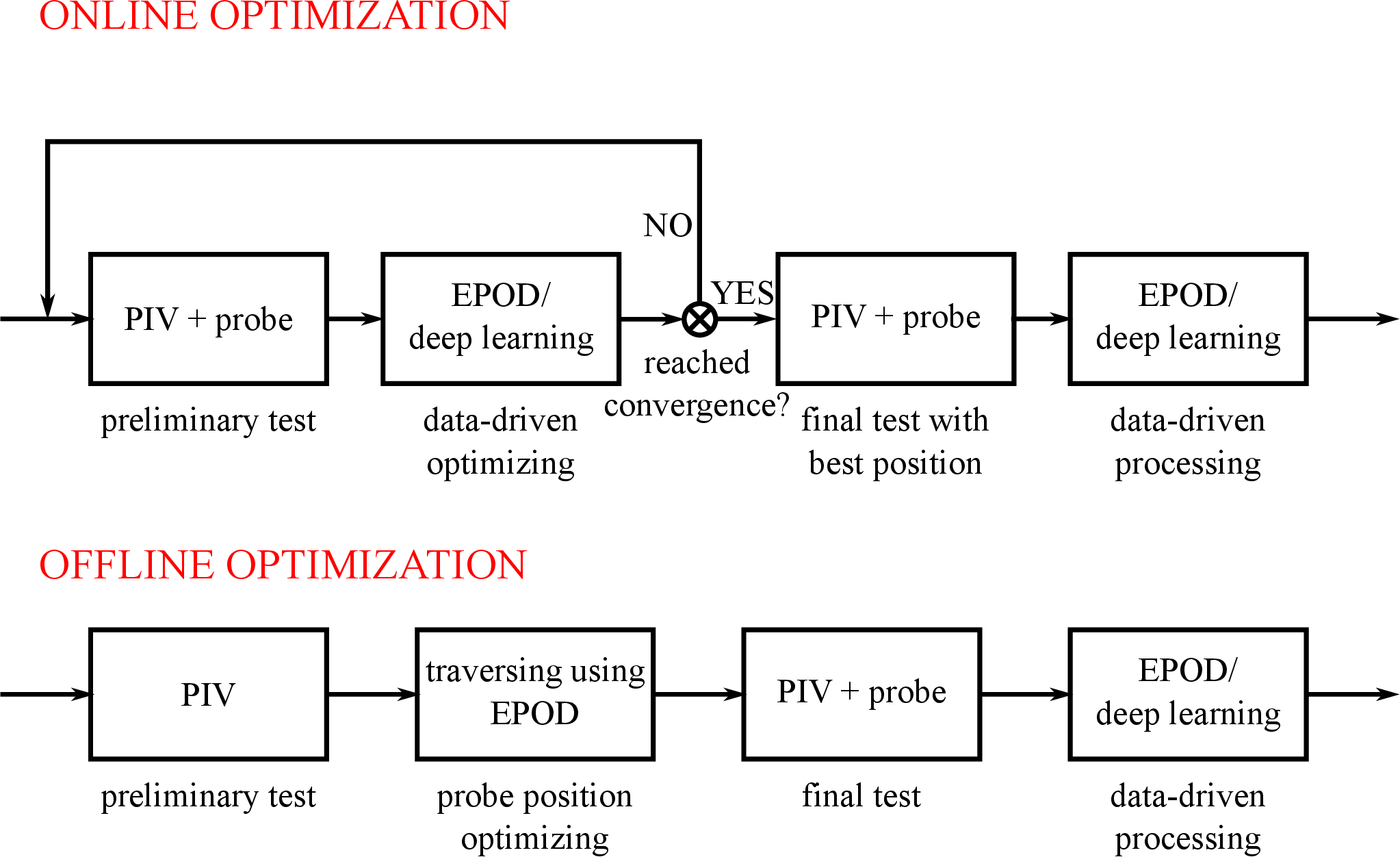}
\caption{The workflow of online (top) and offline (bottom) sensor placement optimization.}
\label{fig: workflow}
\end{figure*}

Data-driven methods, such as the EPOD described in the previous section, can be used to provide temporal resolution to snapshot \ac{PIV} if coupled with high-repetition-rate sensors. An accurate flow estimation allows velocity fields to be obtained ideally at the same temporal resolution as the probes. A sound placement of the sensors is paramount to ensure the quality of the temporal reconstruction of the flow field. Flow estimation includes two phases: i) sensor placement (either based on preliminary tests or on intuition of the user); ii) final testing using both \ac{PIV} and sensors simultaneously to estimate the time-resolved flow field.

The most straightforward method for optimizing sensor placement is to deploy real probes during the first phase and evaluate their performance, a process we refer to as \textbf{``online optimization''}. This approach involves the following steps: 
\begin{enumerate}
    \item placing the probes in the candidate location;
    \item capturing and processing the data from \ac{PIV} and from the probes;
    \item performing the estimation process;
    \item evaluating the quality of estimation and refining the sensor placement until an optimum in the accuracy of the estimation is reached.
\end{enumerate} 
It must be remarked that, even with a small number of probes and limited candidate locations, this approach becomes quickly unaffordable from the practical viewpoint. For $p$ probes and $n$ candidate locations, the number of required \ac{PIV} experiments is $n!/((n-p)!p!)$ for a full coverage of all combinations.

To address this, we propose an \textbf{``offline optimization''} method for sensor placement, relying solely on one dataset of non-time-resolved \ac{PIV} data for the sensor placement, i.e. without requiring physical probes. This approach is sketched in Fig. \ref{fig: workflow}. A similar principle is leveraged in other sensor optimization frameworks based on pattern discovery (for instance in \citealp{manohar2018data}). However, a challenge arises here because delay embedding requires sensor signals or pointwise velocity from time-resolved \ac{PIV}, which snapshot \ac{PIV} cannot provide. The proposed method obviates the need for temporal resolution by using surrogates of the sensor time series extracted from snapshot PIV. In the following section, the nature of these surrogates as well as their difference with real probes time-series will be clarified.

\subsection{Definition of sensor formations}
\label{subsec:sensor_definition}

In advection-dominated flows, sensors are typically placed near the downstream boundary of the domain in order to minimize the (1) wake produced inside the domain, (2) the reflection on the sensors in \ac{PIV} images. In certain applications, placing sensor supports downstream may not be practical, leading to sensors being installed beneath the surface of objects. However, this paper does not consider that specific scenario. As mentioned in Sec. \ref{sec:EPOD}, the probe number can be artificially increased by using time segments of signal. The underlying hypothesis is that, if advection is dominating, the velocity recorded by the probe in time will be similar to the velocity of the flow upstream at earlier time instants.  In this paper, this approach, used by among others \cite{hosseini2015sensor, discetti2018estimation} will be referred to as \textbf{``probes time series''}. This arrangement is possible only by deploying physically the probes in the domain and performing synchronized measurements with \ac{PIV}.

To avoid the time-consuming process of performing multiple probe deployments to identify the optimum sensor placement, we introduce a proxy for physical probe time series by exploiting the temporal and spatial coherence of advection-dominated flows. Specifically, we simulate probe time series using data from an entire row in \ac{PIV} snapshots. These simulated sensors, referred to as ``\textbf{row surrogate sensors}'', are based on the assumption that the flow advects predominantly in the horizontal direction without loss of generality. It is important to note that the row surrogate sensors are usually extracted from single snapshots of non-time-resolved \ac{PIV} fields, thus they are only useful for offline optimization and cannot enhance the temporal resolution of flow fields.

A potential limitation of the row surrogate sensors is that Taylor's hypothesis should be valid throughout at least one convection through-flow time (i.e. the time it takes to the flow to cross the measurement domain). However, it can be expected that this hypothesis generally holds near the probe location, and then becomes less reliable farther upstream. To mitigate this, we propose masking the row sensors based on their correlation with the downstream points of the row (candidate position for the probes). The correlation map $R_{ij,iN}$ between the generic point and the last position in the same row is given by,

\begin{equation}
R_{ij,iN} = \frac{\langle u'_{ij}u'_{iN}\rangle}{\sqrt{\langle u'^2_{ij}\rangle\langle u'^2_{iN}\rangle}}
\label{eqcorr}
\end{equation}

where $u'_{ij}$ is the streamwise fluctuation velocity at the $j$-th position of the $i$-th row and $u'_{iN}$ is the streamwise fluctuation velocity at the downstream end of the same row. The operator $\langle \cdot \rangle$ indicates the temporal average. For planar \ac{PIV}, this equation produces a \ac{2D} distribution, while for volumetric \ac{PIV}, it can be calculated for each cross-section, or averaged across the volume thickness to improve convergence. To mask row sensors, we retain only the points with correlation values above a threshold, ensuring that at least $75\%$ of rows retain more than $75\%$ of their entries. These masked sensors, termed \textbf{``masked row sensors''}, are expected to better simulate probe time series by excluding less correlated regions.

To summarize, three kinds of sensor formation will be considered:
\begin{itemize}
    \item \textbf{Probes time series:} these are the time-series extracted from physical probe measurements (i.e. from probes placed in the flow) as described in Sec \ref{sec:EPOD}. These data would be the ones available during an online optimization. It is the most intensive from the position optimization viewpoint as testing each combination requires a complete experiment. 
    \item \textbf{Row surrogate sensors:} these are the surrogate time-series used by our proposed ``offline'' probe position optimization. It is directly applied to the snapshot \ac{PIV} data by extracting rows of vectors and using them to reconstruct the full field. The combination with the highest reconstruction quality defines the optimal probe location. This method requires 2 experiments: (1) a standard \ac{PIV} experiment for offline optimization, without any probe deployment; (2) an experiment with synchronized \ac{PIV} and probes in the optimal location identified in (1) to perform the time-delay embedding and corresponding flow estimation of time-resolved fields. 
    \item \textbf{Masked row sensors:} these surrogate time-series are analogous to the ones used in offline optimization with row surrogate sensors, the main difference being a masking applied to exclude points with low correlation from the process. The main hypothesis is that points with high correlation within each row would better represent the actual recording of physical probes capturing time series.
\end{itemize}

\subsection{A quick sensor placement method}

The proposed offline sensor placement optimization is based on a brute-force search of possible combinations of rows to obtain the best candidates for flow estimation once the probes are located in the field. This approach is practical if two conditions are met: (1) the estimation process is computationally efficient (e.g., the \ac{EPOD} method used here); (2) the number of candidate locations and sensors is limited. For example, with 20 candidate positions and a set of 3 probes, the number of possible combinations exceeds 1000, which remains computationally manageable.

There are already some method to determine the sensors' position without such brute-force search. We explore an alternative greedy optimization method based on a modified version of QR-pivoting, as described in \cite{manohar2018data}. This method is derived from the QR discrete empirical interpolation method \citep{drmac2016new}.

QR-pivoting is briefly explained here; readers can refer to \cite{manohar2018data} for more details.

Consider a full-field data snapshot represented as a vector $\mathbf{u} \in \mathbb{R}^{1\times n_p}$. For simplicity, we assume only one measured component, though this can be extended to multiple components. If $s$ sensors are available, the sensor measurements $\mathbf{p} \in \mathbb{R}^s$ can be expressed as,

\begin{equation}
    \mathbf{p} = \mathbf{Cu^T}
    \label{S3}
\end{equation}
Here, the measurement matrix $\mathbf{C} \in \mathbb{R}^{s\times n_p}$ consists of zeros and ones, representing whether a sensor is located at that grid point of the velocity field. The field $\mathbf{u}$ can also be expressed as a linear combination of a basis $\mathbf{\Phi} \in \mathbb{R}^{n_p \times n_t}$,

\begin{equation}
    \mathbf{u}^T =\mathbf{\Phi \mathbf{a}}^T,\quad\quad\quad\mathbf{a}\in\mathbb{R}^{1 \times n_t}
    \label{eq:decomposition_vector}
\end{equation}
where $\mathbf{a}$ is a vector of coefficients. As in the previous section, we assume $n_t<n_p$. The basis $\mathbf{\Phi}$ can be predefined (e.g., Fourier modes) or data-driven (e.g., derived from \ac{POD}). If using \ac{POD}, $\mathbf{a}$ contains the time coefficients in any snapshot weighted by singular values. Substituting this into the sensor measurement equation gives

\begin{equation}
    \mathbf{p} = (\mathbf{C\Phi})\mathbf{a}^T
\end{equation}
If the basis includes a number of modes larger than the number of sensors, the system becomes underdetermined. To address this, we reduce the basis to the first $r$ modes, $\mathbf{\Phi}_r \in \mathbb{R}^{n_p \times r}$, which capture most of the variance. This simplifies the equation to

\begin{equation}
    \mathbf{p} \approx (\mathbf{C\Phi}_r)\mathbf{a}_r^T,\quad\quad\quad\mathbf{a}\in\mathbb{R}^{1\times r}
    \label{S5}
\end{equation}
Then the field can be estimated from sensor data $\mathbf{s}_{se}$,

\begin{equation}
    \hat{\mathbf{u}}^T = \mathbf{\Phi}_r\hat{\mathbf{a}}_r^T = \mathbf{\Phi}_r(\mathbf{C\Phi}_r)^\dagger\mathbf{s}_{se}
\end{equation}
where $\hat{\cdot}$ represents estimated value and $^\dagger$ indicates the Moore-Penrose pseudo-inverse.

The QR-pivoting leverages the QR factorization of the spatial mode matrix (reduced to rank $r$) to identify the locations that would minimize the condition ratio of $\mathbf{C\Phi}_r$, thus improving numerical stability. This principle, based on QR discrete empirical interpolation method \citep{drmac2016new}, was used by \cite{manohar2018data}. The matrix $\mathbf{A}$ is decomposed into a unitary matrix $\mathbf{Q}$, an upper triangular matrix $\mathbf{R}$ containing non-increasing diagonal elements, and a column permutation matrix $\mathbf{C}$ such that

\begin{equation}
    \mathbf{AC}^T = \mathbf{QR}
\end{equation}
Here, $\mathbf{A} = \mathbf{\Phi}_r$ when $p = r$, and $\mathbf{A} = \mathbf{\Phi}_r \mathbf{\Phi}_r^T$ when $p > r$. The QR factorization is performed column by column using the Householder transformation, selecting columns that maximize stability at each step.

Due to factors such as inhomogeneous illumination or particles moving into/out of the \ac{FOV}, \ac{PIV} accuracy often deteriorates near the boundaries. This limitation makes applying QR-pivoting only with data near the downstream edge less practical. \newline
In this work, we modify the QR-pivoting approach to handle multiple signal entries from each sensor. Instead of selecting individual sensor locations, we consider entire rows of \ac{PIV} data as surrogates for the time series captured by probes. This requires a modification to pivoted QR decomposition referred to as ``block-pivoted QR'' in this paper.

The block-pivoted QR algorithm uses the same principles as QR-pivoting but evaluates blocks of data (entire rows in the flow field) instead of individual columns (single points in the flow field). This ensures that the time-delay structure is preserved while identifying the best sensor placement for flow estimation.

\vspace{0.3cm}

\section{Validation}
\label{sec:validation}

The proposed offline optimization method is assessed with both synthetic and experimental data. Two datasets from \ac{DNS} and one from experiments are utilized. The details are reported in the remainder of this section.

From this point on, the performance of the techniques used for sensor placement is tested by the velocity reconstruction error using \ac{EPOD}. During testing, we temporally downsample the velocity field to simulate non-time-resolved PIV data, while use the signal located at several point of the velocity field as the sensor data, then the result is compared to the reference time-resolved field. The error $\epsilon$ is defined by
\begin{equation}
    \epsilon = \frac{\|\mathbf{U}_{est}-\mathbf{U}_{ref}\|_2}{n_{test}\times n_cn_p} \frac{1}{U_0}
    \label{eq: error}
\end{equation}

\noindent where the \ac{EPOD} estimated velocity field $\mathbf{U}_{est}$ and reference velocity field $\mathbf{U}_{ref}$ from the simulation or experiment are matrices of the size $n_{test}\times n_cn_p$, $n_{test}$ is the number of snapshots used for testing, and the definition of $n_c$ and $n_p$ follows Sec. \ref{sec:EPOD}. The $\|\mathbf{U}\|_2$ is an entry-wise Frobenius norm for an $n_{test}\times n_cn_p$ matrix $\mathbf{U}$ , which follows
\begin{equation}
    \|\mathbf{U}\|_2 = \left(\sum_{i=1}^{n_{test}}\sum_{j=1}i^{n_cn_p}|u_{ij}|^2\right)^{1/2}
\end{equation}
where $u_{ij}$ is one of the entries of $\mathbf{U}$. Finally, the error is normalized by the bulk velocity or free stream velocity $U_0$.

The optimal sensor positions from probes time series, row surrogate sensors, row sensors masked, block-pivoted QR in Sec. \ref{sec:methods} as well as equidistant sensor placement will be discussed in this paper on the result from the cases below.

\subsection{Fluidic pinball}
The first dataset is from a \ac{2D} \ac{DNS} of the wake of a \textit{fluidic pinball} \citep{deng2020low}. The pinball setup consists of three cylinders with diameter $D$, arranged in an equilateral triangle configuration with a side length of $1.5D$. The triangle is oriented with one vertex pointing upstream. The Reynolds number is set equal to $130$, based on the freestream velocity and cylinder diameter. These conditions fall within the chaotic regime, as illustrated by \cite{deng2020low}.

The flow field extends from $x = 1D$ to $x = 7D$ and from $y = -3D$ to $y = 3D$, where $x$ and $y$ denote the streamwise and crosswise directions respectively. The velocity data from the original \ac{DNS} mesh has been interpolated onto a Cartesian grid with a spacing of $0.08D$, producing $76\times76$ vector fields to simulate the \ac{PIV} fields.

All combinations of three probes time series, row surrogate sensors, or masked row surrogate sensors are tested. A total of $26$ potential positions, uniformly spaced $0.24D$ apart in the $y$-direction, is explored. This results in $2600$ combinations. The probe data only considers streamwise component of velocity. When simulating flow estimation with time-delay embedding, the time segment is set to $80$ steps with a time spacing of $0.08D$, thus resulting in a horizon slightly larger than the convection time through the domain. The dataset is split into training ($4685$ non-time-resolved snapshots) and testing ($4900$ time-resolved frames).

\subsection{Turbulent channel flow}
A more challenging test case is based on the \ac{DNS} of a turbulent channel flow from the Johns Hopkins Turbulence Databases. This dataset is characterized by higher spectral richness, making the \ac{EPOD} estimation more difficult compared to the shedding-dominated wake flow of the previous pinball test case. The \ac{DNS} for this dataset is solved in a domain of size $8\pi h\times2h\times3\pi h$, where $h$ is the half-channel height, with a resolution of $2048\times512\times1536$ nodes, and at a Reynolds number $Re_\tau\approx 1000$, given the bulk velocity $U_b=0.99994$, friction velocity $u_{\tau}=4.9968\times10^{-2}$ and viscosity $\nu=5\times10^{-5}$. The reader is referred to \cite{li2008JHTDB} for further details on the simulation.

Similarly to \cite{discetti2018estimation}, sub-domains extending from the wall to the centreline of the channel, and with a streamwise length of $h$, are extracted from the full domain. The advantage of this approach is that the dataset size can be increased by exploiting statistical homogeneity in the spanwise and streamwise direction. 

We are going to consider four different scenarios: (1) standard \ac{2C} planar \ac{PIV}; (2) two-dimension-three-component (2D-3C) stereoscopic \ac{PIV}; (3) volumetric \ac{3C} measurements in a volume with aspect ratio, representing the ratio of the volume depth to $h$, of 1/8 and (4) 1/2 to simulate tomographic \ac{PIV}. The \ac{2D} sub-domains include $88\times88$ velocity vectors with a grid spacing of $0.0114h$, and the \ac{3D} sub-domains have an aspect ratio from $0$ to $1/2$. The training data are independently sampled in space and time to minimize correlation between each other. Leveraging translational symmetry, this study assumes that the streamwise coordinate spans only between $0$ and $h$, although the actual domains considered are located at different streamwise positions. The $y$ direction ranges between $0$ and $h$, with $0$ being the position of the wall.

The probe series contains only the streamwise component of velocity. The time spacing for probe data with time series is $0.0065$, and segments of $152$ samples are taken at each location, thereby covering the convection throughtime of one subdomain length. The $30$ potential probe positions are uniformly distributed along the downstream edge of the domain for the cases of \ac{2D} fields. For the case of 3D domain, the probes are placed along the downstream edge of the mid-plane in the spanwise direction, with the same spacing as the 2D case. Considering that 3 probes are used for the reconstruction, this results in $4060$ combinations. $7200$ non-time-solved snapshots are used in training and $3600$ time-resolved frames are used for testing for all scenarios.

\subsection{3D experiment of propeller wake}

\begin{figure}
\centering
\hspace{-8mm}
\includegraphics[scale=0.15,trim=32mm 0mm 32mm 0mm,clip]{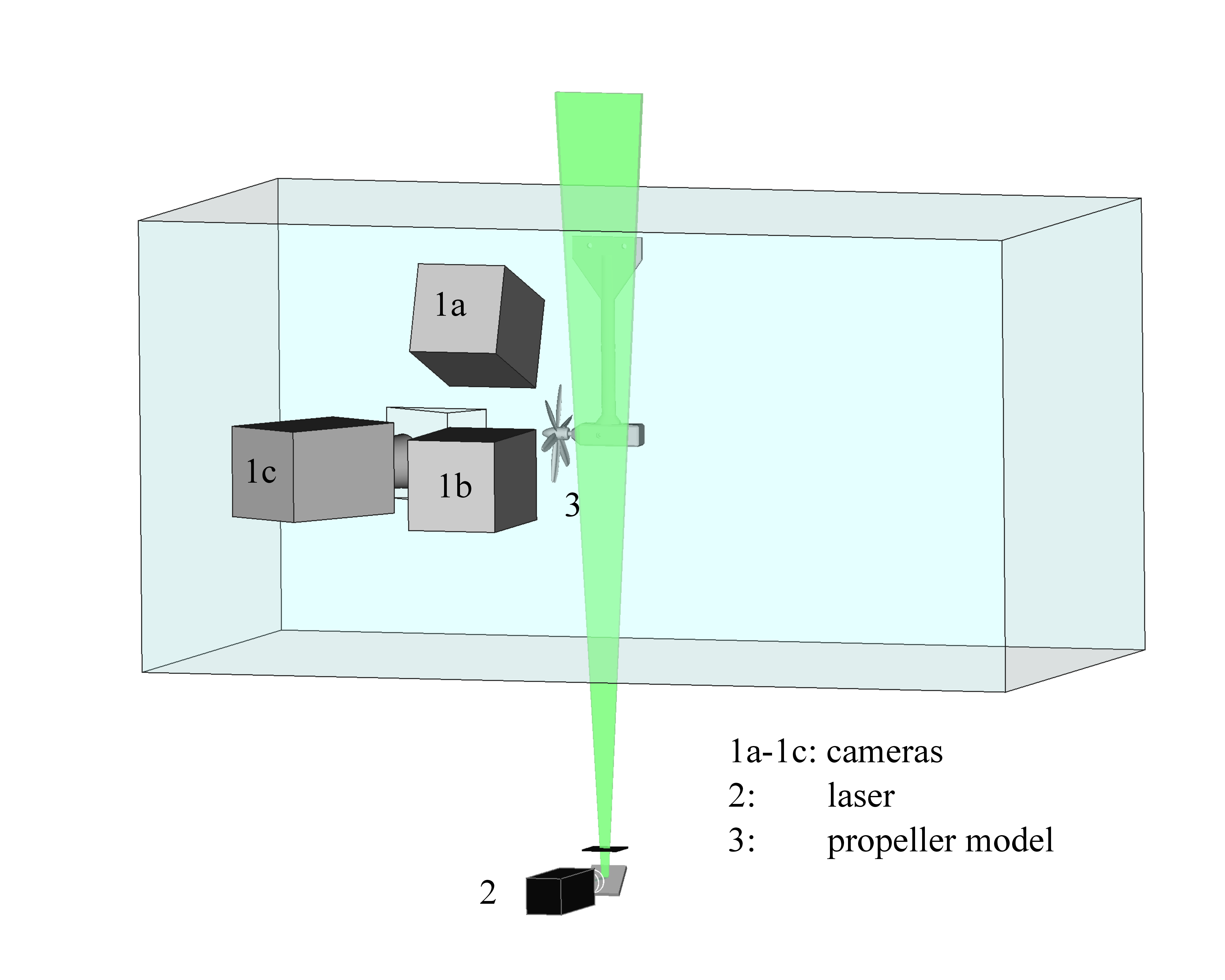}
\caption{Sketch of the experimental setup of tomographic \ac{PIV} measurement in the wake of a windmilling propeller.}
\label{fig: blades_setup}
\end{figure}

The sensor positioning is also validated in a 3D experimental case in the wake of a windmilling propeller model. The experiment was carried out in the water tunnel at Universidad Carlos III de Madrid. A 3D printed powerless 6-blade propeller with a hub-to-tip radius $r=55 \,\unit{mm}$ was installed on a pylon with null angle-of-attack with respect to the free stream, as shown in Fig. \ref{fig: blades_setup}. The blades are driven by the water flow at a far-field velocity $U_{\infty}=0.15 \,\unit{m/s}$, reaching a rotation speed of about $6.0\, \,\unit{RPM}$. The Reynolds number and the Strouhal number based on the diameter are $16500$ and $0.22$, respectively.

A time-resolved thin tomographic \ac{PIV} is performed in the wake region, covering a volume of $90\times90\times7\,\unit{mm^3}$, at a distance of $20 \,\unit{mm}$ from the propeller axis. The flow is seeded with polyamide particles with $56\,\unit{\mu m}$ diameter. Three Andor sCMOS cameras, each with $5.5\, \text{Mpixels}$, are deployed and operated at $80 \,\unit{Hz}$ after sensor cropping. A $3\,\unit{W}$ green continuous laser pointer is used for illumination. The exposure time of the cameras is set to $2 \,\unit{ms}$. The cameras are equipped with $50 \,\unit{mm}$ objectives, with $0.5\times$ fisheye lens. The $f_\#$ is set to 8 for the camera in forward scatter, and to 5.6 for the other cameras, to obtain similar imaging conditions.  

The particle images are preprocessed with a sliding-minimum subtraction on a $7\times 7$ pixel kernel and Gaussian smoothing on a $3\times 3$ pixel kernel with a standard deviation equal to 0.75. The images were then
processed using a custom-made tomographic PIV software from Universita' di Napoli Federico II \citep{discetti2013spatial}. First, a volumetric self-calibration \citep{wieneke2008volume} is carried out to reduce the calibration error. The volumetric reconstruction is carried out with a multi-resolution process \citep{discetti2012fastMR}. The reconstruction algorithm is based on the camera-simultaneous multiplicative algebraic reconstruction technique (cSMART). The cSMART is a modified version of the SMART procedure proposed by \cite{atkinson2009efficient} which uses the cameras sequentially. The process consists of 3 CSMART iterations on a 2x binned configuration, 3 CSMART iterations and further 3 SMART iterations on the final resolution of $11\, \text{voxels/mm}$.  The reconstructed distributions are then interrogated with a multi-pass 3D cross-correlation \citep{elsinga2006tomographic} based on direct sparse correlations \citep{discetti2012fastCC}. The final interrogation spot is set to $32^3\,\text{voxels}$, with an overlap of $75\%$. The vector spacing in each direction is $727 \,\unit{\mu m}$.

A non-time-resolved dataset is generated by randomly selecting $4800$ snapshots among the $13620$ snapshots of the original sequence, and $500$ frames are used for testing. The associated probe data are extracted from the full data sequence at the original sample rate from the downstream part of the \ac{PIV} region on the middle plane of the volume span. The portion of tomographic \ac{PIV} snapshots used for flow estimation contains $85\times50\times7$ vectors. Similarly to the synthetic\sout{ test} case of the channel flow, $85\times50$ subdomains of \ac{2C} and 3C velocity fields are extracted for testing to simulate planar \ac{PIV} and stereoscopic \ac{PIV} configurations.

\FloatBarrier

\section{Results}
\label{sec:results}

\subsection{Fluidic Pinball}

\begin{figure}
\centering
\hspace{-12mm}
\includegraphics[width=0.65\linewidth,trim=2mm 0mm 16mm 0mm,clip]{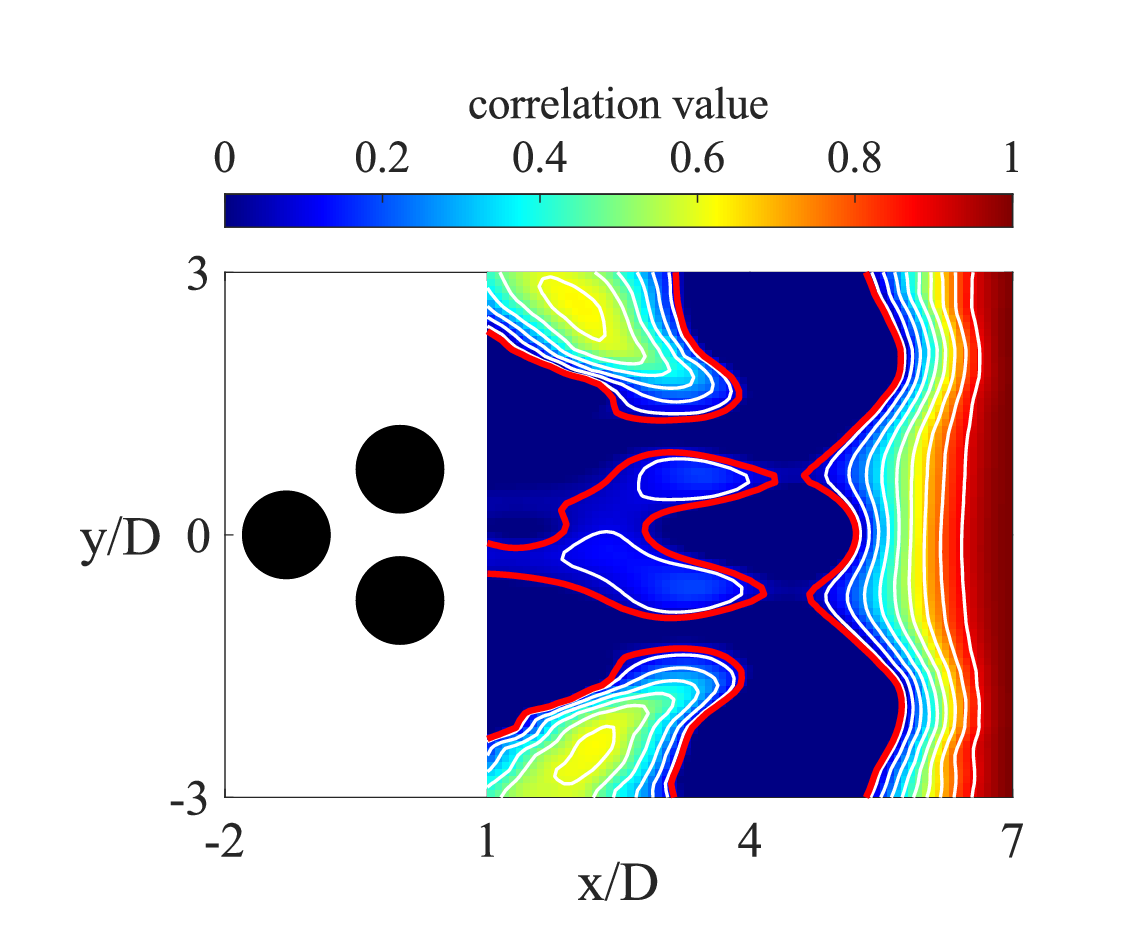}
\caption{Map obtained by stacking the correlation of velocity in each row with the most donwstream point of it, used for masking on the dataset of the pinball. The thick red curves bound the masked region due to a weak correlation.}
\label{fig: pinball_cmap}
\end{figure}

The offline optimization process is assessed first for the case of the fluidic pinball. As discussed in Sec. \ref{sec:methods}, the map obtained by stacking the correlation of row data with their respective downstream edge point (as in Eq. \ref{eqcorr}), is shown in Fig. \ref{fig: pinball_cmap}. The thick red line locates the masking threshold. The region with the correlation value under the threshold is blanked for the case of estimation with masked row sensors. 

\begin{figure}
\centering
\includegraphics[width=0.7\linewidth,trim=2mm 0mm 2mm 8mm,clip]{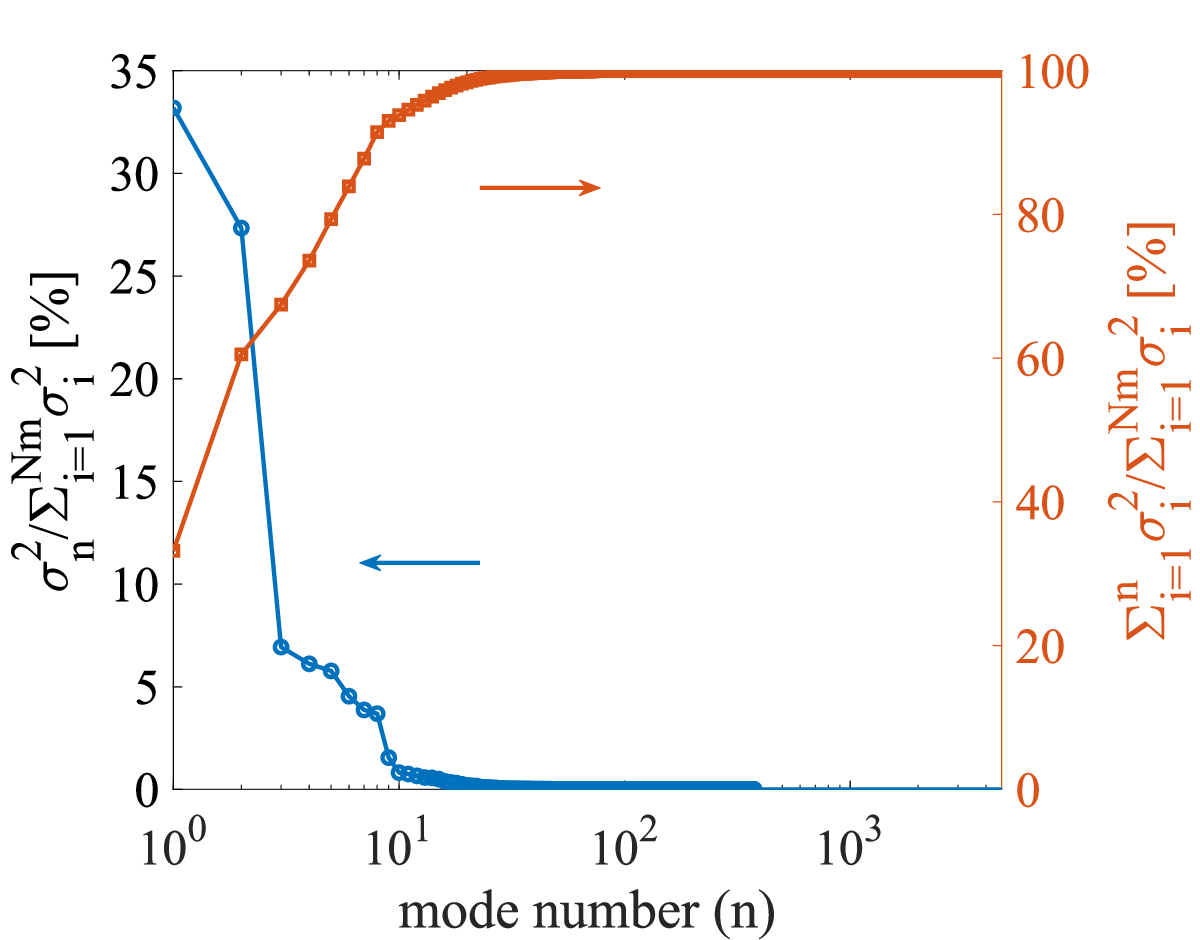}
\caption{The POD spectrum for the pinball dataset. The blue line shows the squared singular values $\sigma_i^2$. The orange line corresponds to the cumulative sum of $\sigma_i^2$. In both cases, the data are normalized with the sum of all the $\sigma_i^2$.}
\label{fig: pinball_spectrum}
\end{figure}

The POD spectrum is presented in Fig. \ref{fig: pinball_spectrum}. The spectrum reports the squared diagonal elements $\sigma_i^2$ of the diagonal matrix $\mathbf{\Sigma}$ in Eq. (\ref{eqn1}), which are related to the variance content corresponding to the 
$i$-th mode \citep{berkooz1993proper}.  The figure shows that the POD spectrum is quite compact, as expected for a low-Reynolds-number shedding-dominated flow. The first two modes contain already $60\%$ of the energy and correspond to the vortex shedding. Ten modes are sufficient to reconstruct $90\%$  of the energy. For this test case, it is expected that a small number of sensors should be able to achieve reasonably good reconstruction accuracy.

\begin{figure}
    \centering
    \includegraphics[width=0.7\linewidth,trim=16mm 0mm 16mm 4mm,clip]{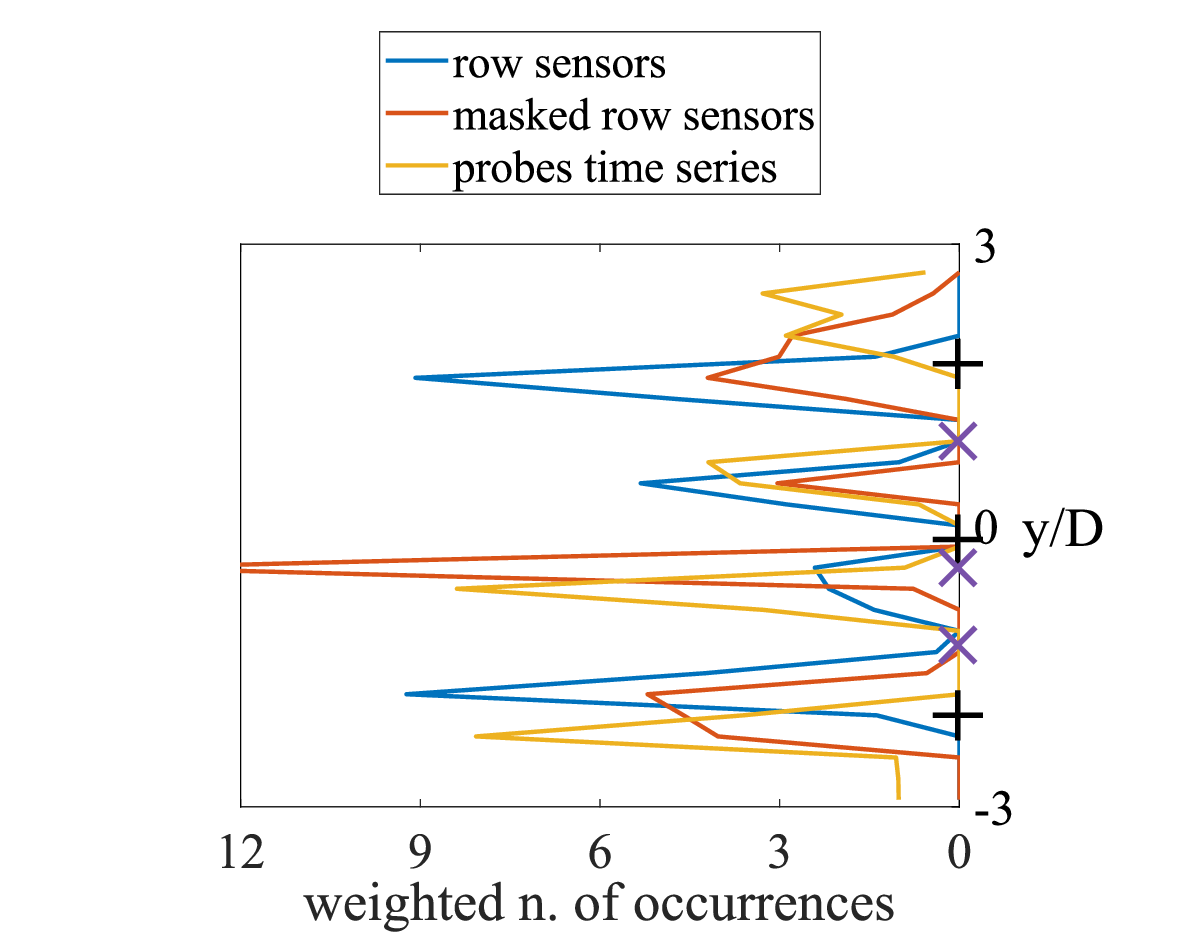}
\caption{The one-dimensional weighted positional frequency distribution of the pinball case, with the definition of $W$ from Eq. \ref{eq:weight}. The symbols $+$ and $\times$ represent the sensor positioning from equidistant sensors and block-pivoted QR respectively.}
\label{fig: pinball_hmap}
\end{figure}

The positional frequency distribution on Fig. \ref{fig: pinball_hmap} represents the best $24$ sensor position combinations of $3$ probes, according to the reconstruction error defined by Eq. \ref{eq: error}. To give more relevance to the most performing positioning, the combinations are sorted according to their reconstruction accuracy, and their contribution to the positional frequency distribution is weighted. The weights are set according to the following relation:
\begin{equation}
    W = e^{\textstyle-\frac{i-1}{N-1}}
    \label{eq:weight}
\end{equation}
with $i$ representing the rank of the sensor position combination after sorting by the error value, and $N$ the number of combinations used to generate the distribution ($24$ in all the cases used in this paper). Four preferred locations were identified from all types of sensors: two symmetrically located outside the wake and two on the edge of the wake. The positional distribution of row sensors is reasonably well aligned with that of the probes time series, even though the two outmost peaks are slightly shifted inwards. This suggests that the row sensors proposed for the offline optimization are reasonably good surrogates for the placement of the physical probes. Additionally, further improvement seems to be achieved by the masking, since the peaks from masked row sensors lay closer to those from point probes time series. 

The figure also includes the best combination according to the block-pivoted QR positioning. The selected positions are mostly concentrated towards the inner part of the wake, thus missing mapping the outer regions. It can be expected that this greedy optimization method will be less performing when reconstructing the fields.

\begin{figure}
\centering
\hspace{14mm}
\begin{subfigure}
    \centering
    \includegraphics[width=0.65\linewidth,trim=10mm 100mm 10mm 0mm,clip]{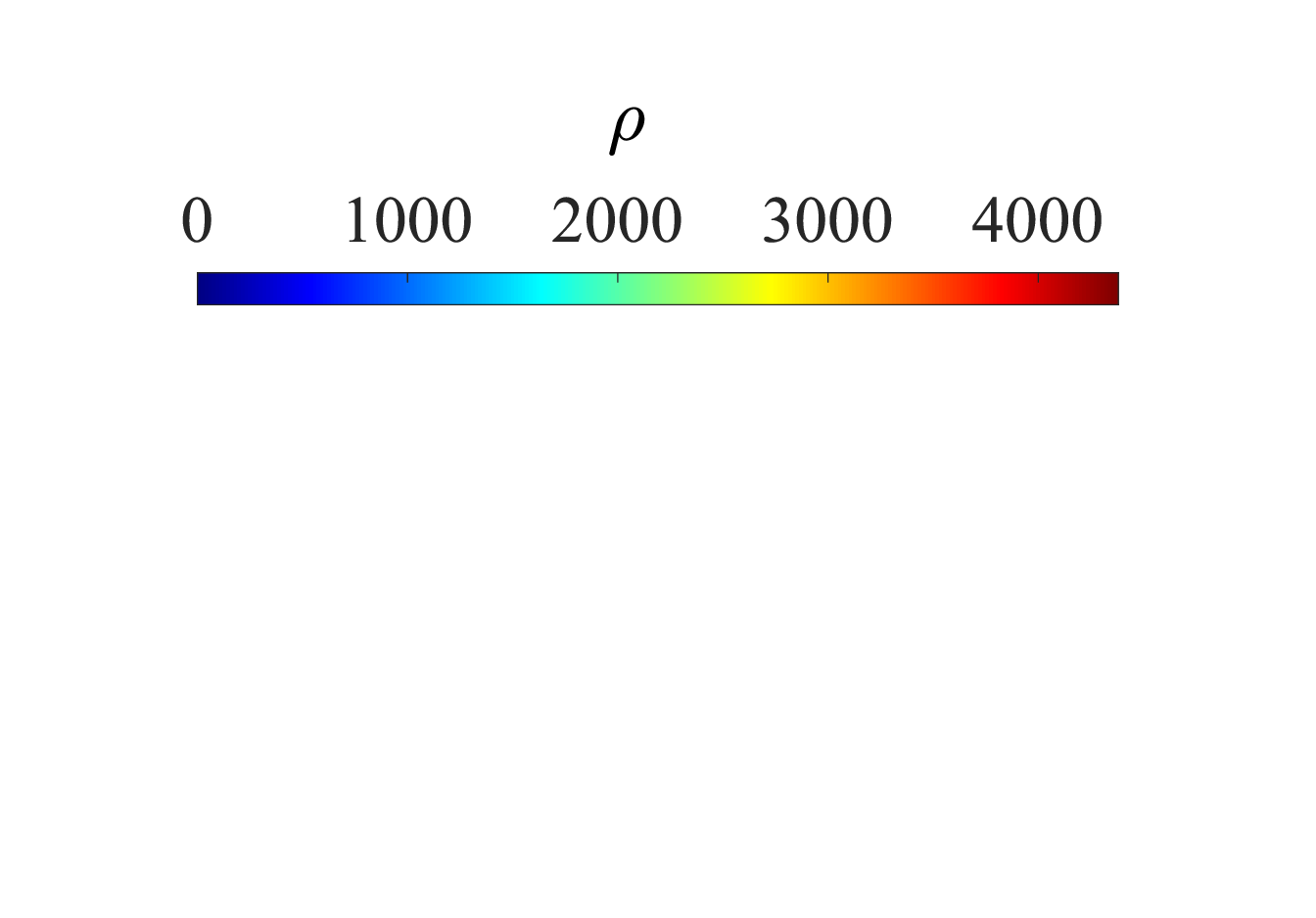}
\end{subfigure}
\newline
\setcounter{subfigure}{0}
\begin{subfigure}[]
    \centering
    \includegraphics[width=0.75\linewidth,trim=0mm 0mm 12mm 0mm,clip]{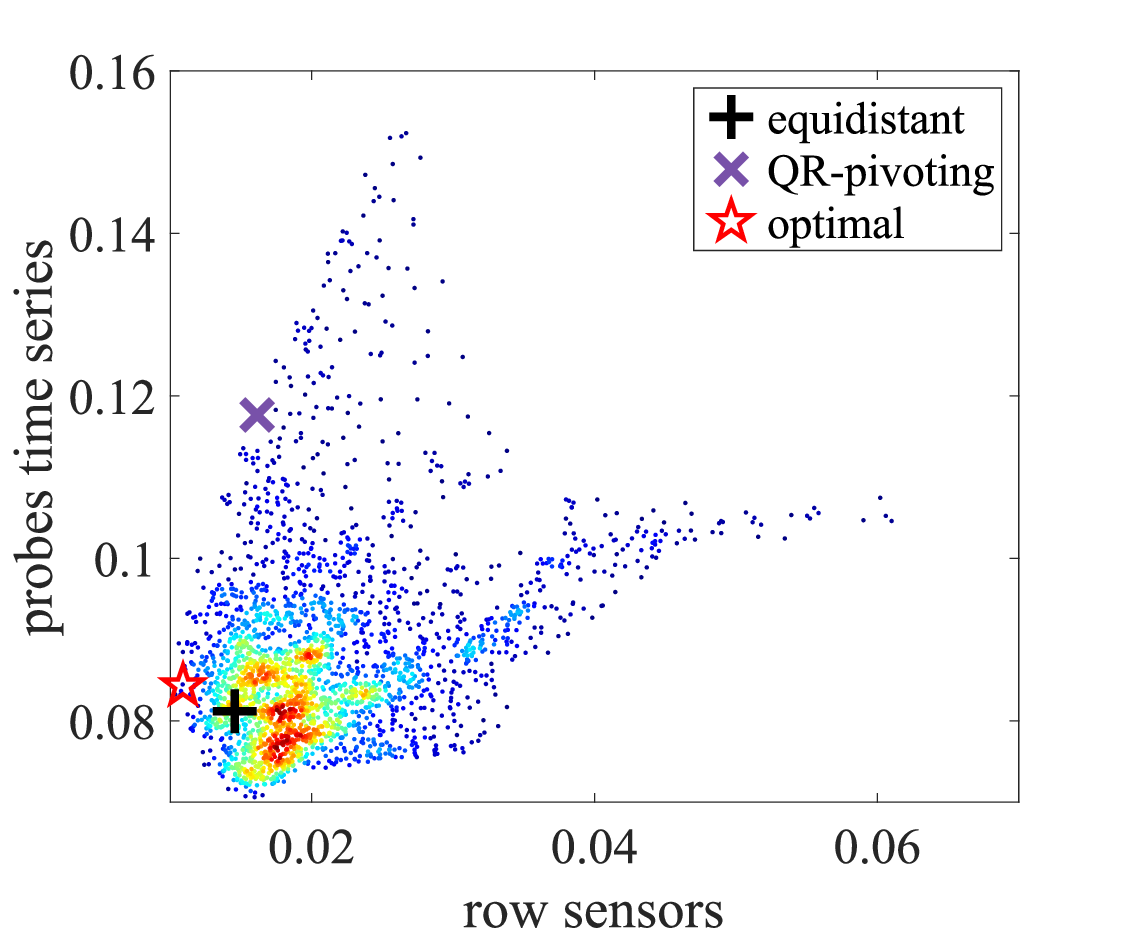}
\end{subfigure}
\hspace{6mm}
\begin{subfigure}[]
    \centering
    \includegraphics[width=0.75\linewidth,trim=0mm 0mm 12mm 0mm,clip]{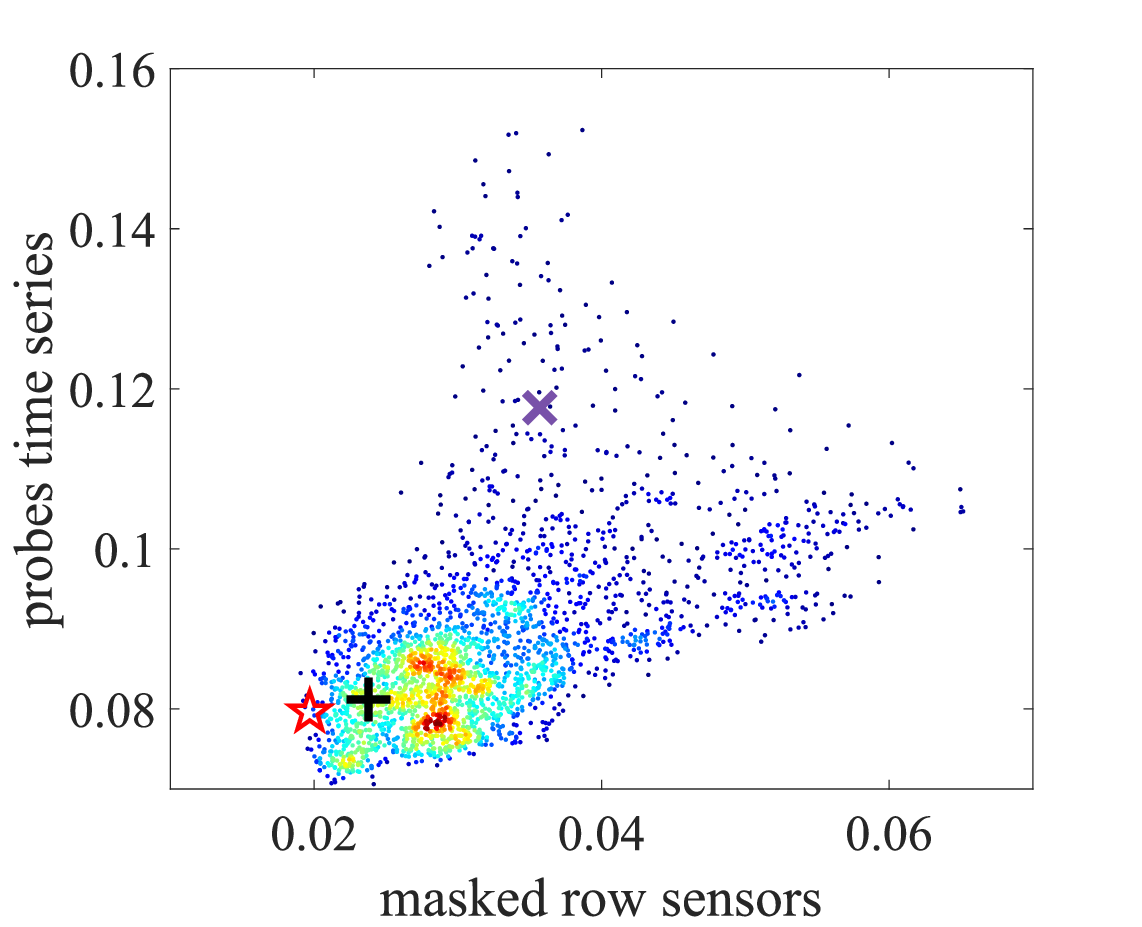}
\end{subfigure}
\caption{The 2D probability density function of the velocity reconstruction error for probes time series versus (a) row sensors and (b) masked row sensors. The colours in the scatter plots serve as auxiliary markers, indicating the local data density. The density is defined as $\displaystyle\sum_i\rho\Delta s_i = 1$, where the $\Delta s_i$ is any local area on the scatter plots.}
\label{fig: pinball_scatter}
\end{figure}

The \ac{2D} distribution of the estimation error, as defined in Eq. \ref{eq: error}, is shown in Fig. \ref{fig: pinball_scatter}. Compared to the {1D} histogram, the \ac{2D} scatter map more effectively illustrates the correlation between different sensor formations. In these scatter maps, each spot represents a specific combination of sensor locations, with the x-axis and y-axis indicating the estimation error. The likelihood of data corresponding to specific x and y positions could be inferred from the density of points in those areas. To help readability, the scatter plot has been colour-coded for the probability density function.
Here, the density $\rho$ is counted by dividing the whole map into $100\times100$ regions and counting the number of spots, then normalized by the total number of spots thus $\displaystyle\sum_i\rho\Delta s_i = 1$ is complied, where the area $\Delta s_i$ follows the unit of x and y axis. For the case of row surrogate sensors (with and without masking), the error corresponds to the reconstruction error using directly the row data from each snapshot. When surrogate sensors are replaced with probes measuring time series and time-delay embedding is used in the reconstruction, the error is expected to be larger due to the lower correlation between the flow field snapshot and the probes time series.

A good qualitative indicator that offline optimization with rows is a good surrogate for sensor placement of the direct measurements is the shape of the error map. A stretched linear shape indicates that the two error metrics are interchangeable, thus the search for optimal placement can also be carried out with the row sensor. Ideally, if the leftmost point of the histogram corresponds also the downmost point, it would mean that the analysis of the reconstruction error to row sensors is able to identify also the best combination for flow estimation with probes time series.

In the case of row sensors without masking, this happens to a reasonable extent. We consider the positions obtained by selecting the three peaks of the positional frequency distribution and interpolating them with a spline to refine the positions. The positioning we obtain, corresponding to the leftmost point of the scatter plot, has an error about $19\%$ higher than the effective optimal placement based on probes time series, as given in Table \ref{tab: pinball_error}. The block-pivoted QR provides worse performances, as expected by Fig. \ref{fig: pinball_hmap}. The concentration of probes inwards in the wake region does not provide enough details of the edge of the wake for the reconstruction. 

Equidistant probe placement (thus blind to the flow distribution) works already with nearly optimal performances. This might be ascribed to the small size of the domain in the spanwise direction, in which equidistant probe spacing allows easy capture of both the shedding structures and the wake meandering. 
The best performance is provided by the masked row sensors, which provide an error only $13\%$ higher than the optimal combination of probes with time series. This is achieved using solely the snapshot \ac{PIV} data, while optimization with probes time series would have required a large number of tests with probes and fields simultaneously. While this was possible in simulations, in experiments this would have been extremely onerous. 

The scatter plots confirm that masked row sensors have a higher correlation to probes time series, better representing sensor locations. A sharper down-left corner in the scatter plot indicates higher correlation among preferred locations with the least velocity reconstruction error.

\begin{table*}[]
\centering
\begin{tabu}{|[2pt]l|[2pt]c c c c|[2pt]c|[2pt]}
    \tabucline[2pt]{-}
    \multirow{2}{*}{\diagbox{field type}{probe type}} & \bf{equidistant} & \bf{block-} & \bf{surrogate} & \bf{surrogate masked} & \bf{probes} \\ 
    & \bf{probes} & \bf{pivoted QR} & \bf{row sensors} & \bf{row sensors} & \bf{time series} \\
    \tabucline[2pt]{-}
    \bf{planar PIV} & 0.0812 & 0.1176 & 0.0843 & 0.0796 & 0.0706 \\
    \tabucline[2pt]{-}
\end{tabu}

\caption{The error of velocity reconstruction using time-delay embedding and sensors located according to different methods for the fluidic pinball. The error is defined as in Eq. \ref{eq: error}. In the last column, the optimal case of online optimization with probes time series is included for reference.}
\label{tab: pinball_error}
\end{table*}

\subsection{Turbulent channel flow}

\begin{figure}
\centering
\begin{subfigure}[]
    \centering
    \includegraphics[width=0.6\linewidth,trim=4mm 16mm 4mm 0mm,clip]{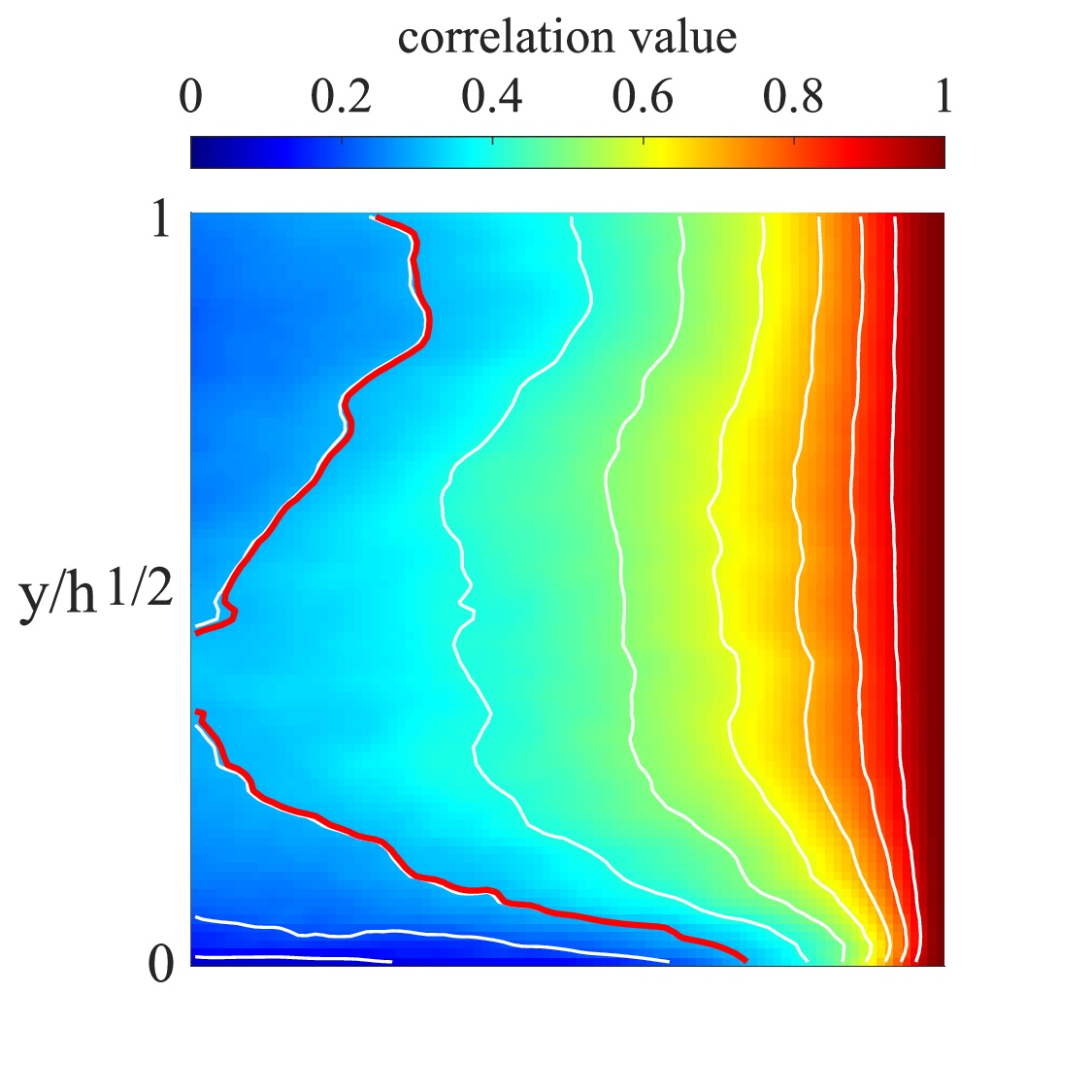}
\end{subfigure}
\begin{subfigure}[]
    \centering
    \includegraphics[width=0.6\linewidth,trim=4mm 18mm 4mm 18mm,clip]{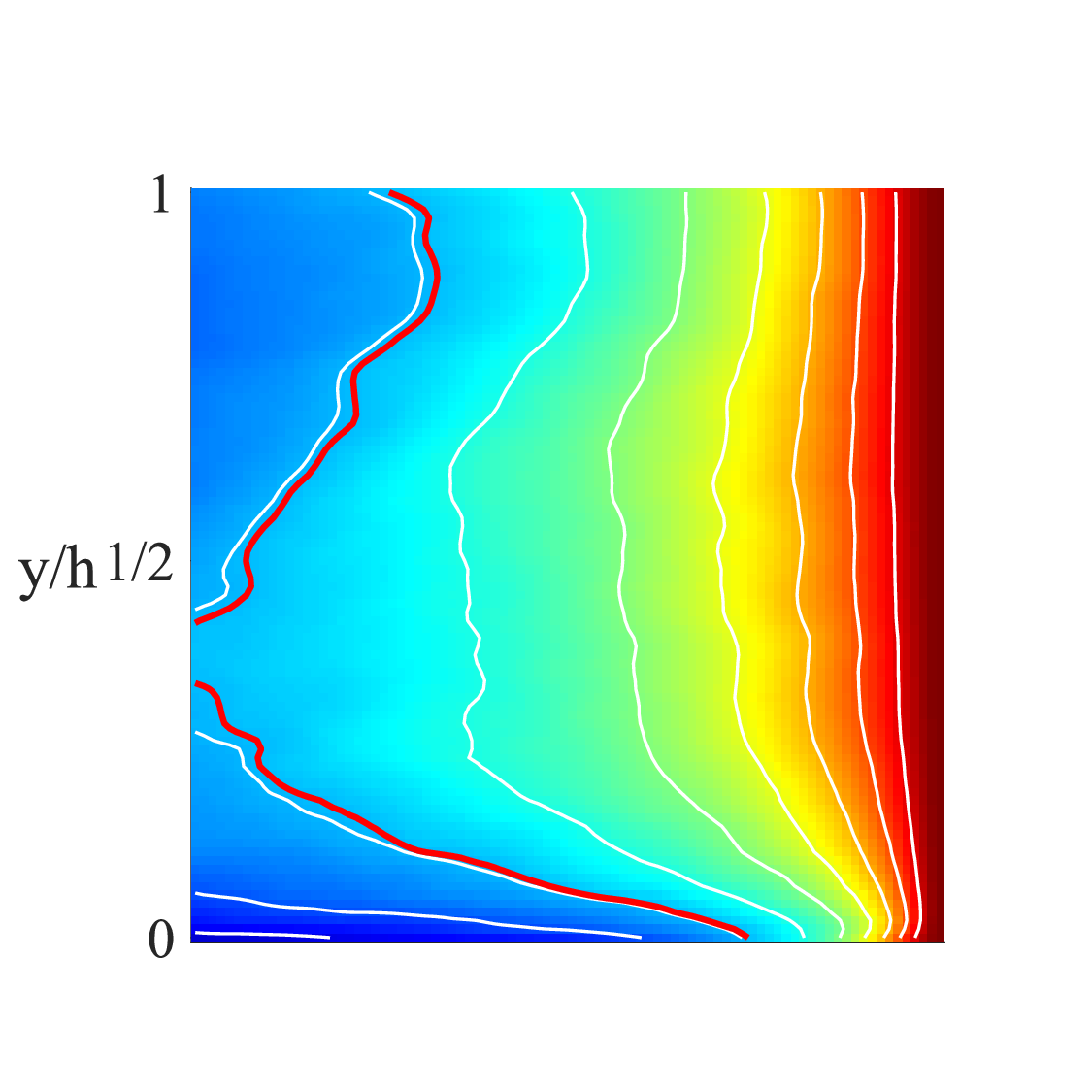}
\end{subfigure}
\begin{subfigure}[]
    \centering
    \includegraphics[width=0.6\linewidth,trim=4mm 0mm 4mm 16mm,clip]{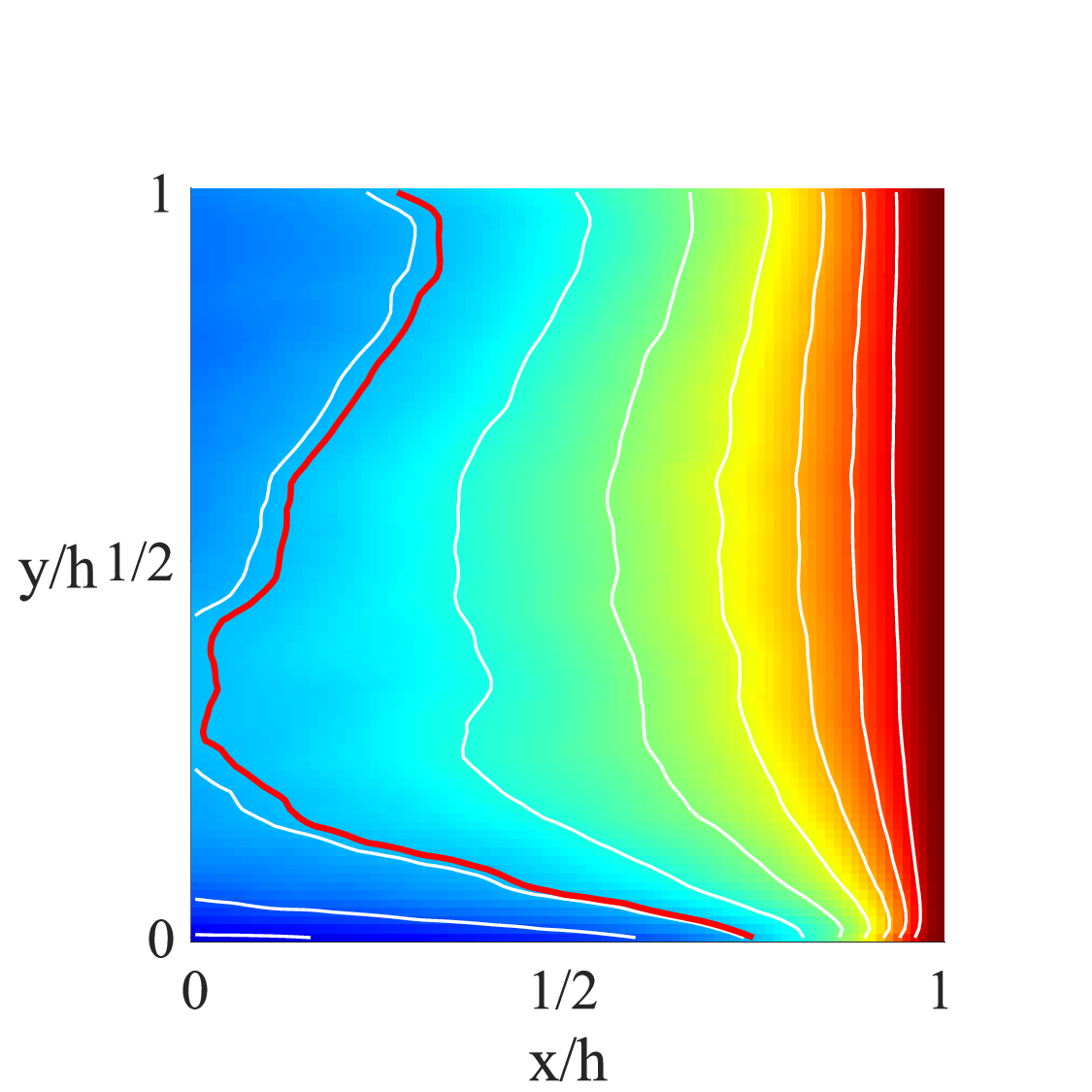}
\end{subfigure}
\caption{Map obtained by stacking the correlation of velocity in each row with the most donwstream point of it, used to mask the two sensors in the turbulent channel flow, with different aspect ratio: (a) 2D plane, (b) 1/8 and (c) 1/2. The thick red curves bound the masked region due to a weak correlation.}
\label{fig: channel_cmap}
\end{figure}

The turbulent channel flow presents the challenge of higher spectral richness. On the other hand, since Taylor's hypothesis is expected to be valid apart from the near-wall region, it might be argued that the row surrogate sensors should be better representative of the probe time series. The correlation maps for the channel case are depicted in Fig. \ref{fig: channel_cmap}. Subfigure (a) represents the correlation maps for both the 2D-\ac{2C} and 2D-3C fields to simulate planar and stereoscopic \ac{PIV}, as they share the same map. Subfigures (b) and (c) represent the map of thin and thick volumetric fields. In all cases, we observe a nearly monotonic decrease in correlation from downstream to upstream (in the $-x$-direction). The wall distance has a minor effect on the correlation in the region closer to the probe. On the other hand, towards the upstream edge, a stronger correlation is observed in the region ranging from $y/h = 0$ to $y/h = 0.4$, followed by a gradual decrease until $y/h = 1$. This trend is consistent across all correlation maps. The lower correlation values near the wall are likely due to the lower convective velocity and strong shear, which cause significant flow distortion before it reaches the downstream edge, as well as 3D motion. A slightly lower correlation in the channel centre might be explained by the interaction with the flow on the opposite side of the channel.

\begin{figure}
\centering
\begin{subfigure}[]
    \centering
    \includegraphics[width=0.48\linewidth,trim=2mm 20mm 28mm 0mm,clip]{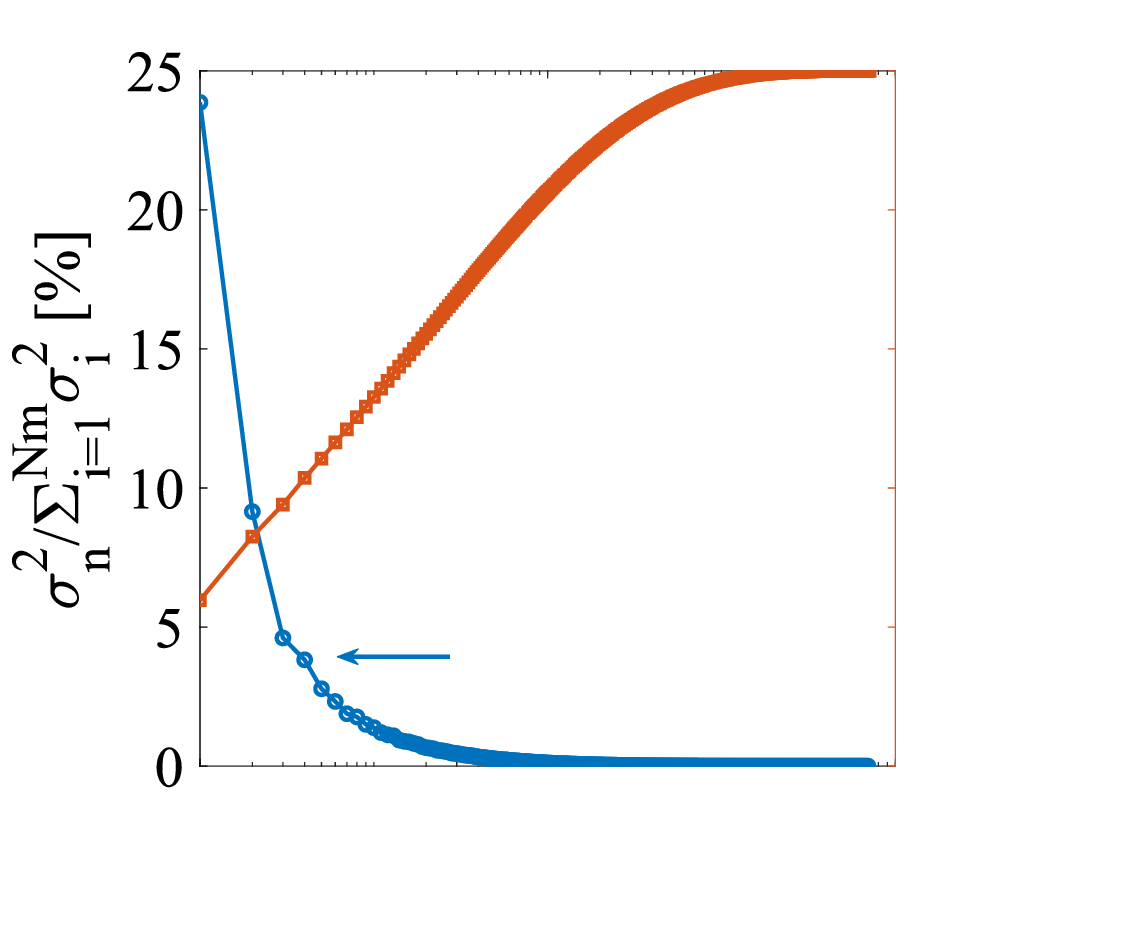}
\end{subfigure}
\hspace{-2mm}
\begin{subfigure}[]
    \centering
    \includegraphics[width=0.48\linewidth,trim=28mm 20mm 2mm 0mm,clip]{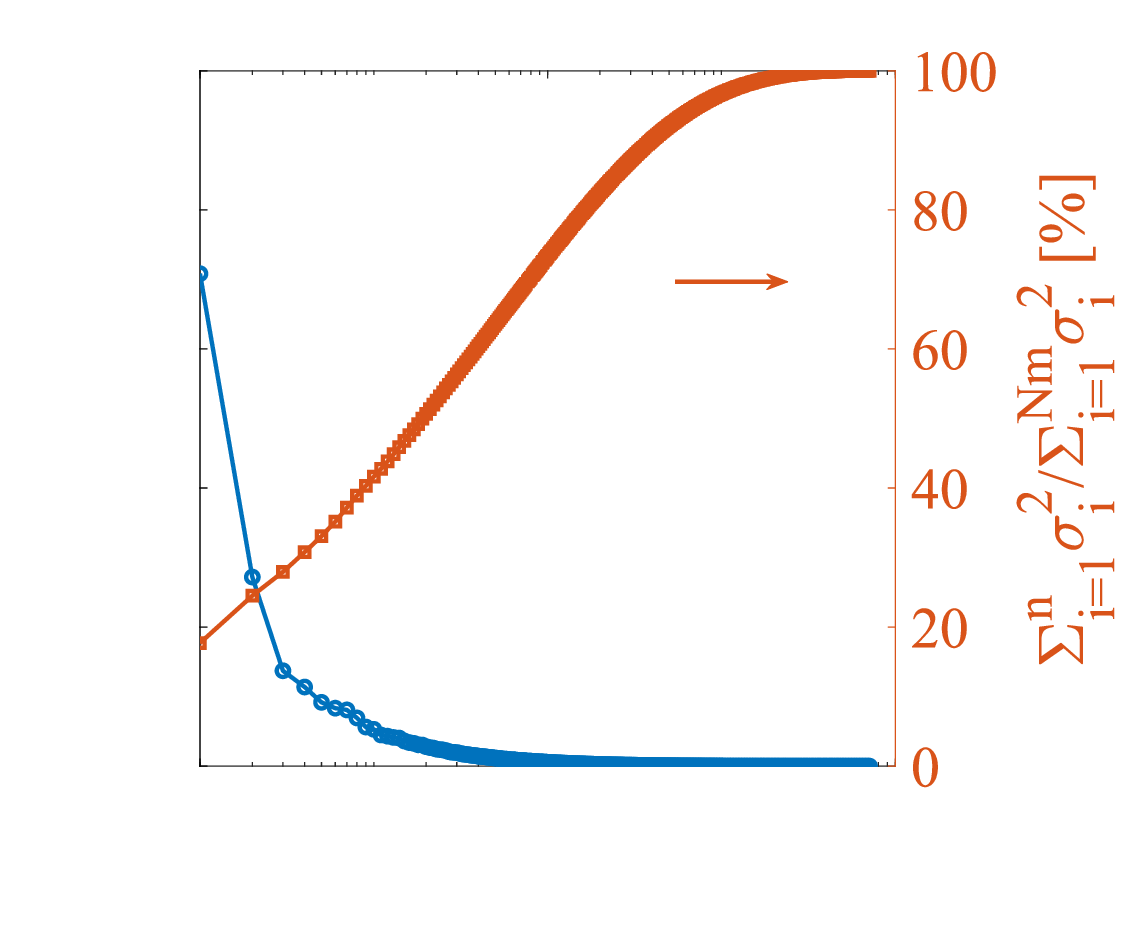}
\end{subfigure}
\begin{subfigure}[]
    \centering
    \includegraphics[width=0.48\linewidth,trim=2mm 0mm 28mm 0mm,clip]{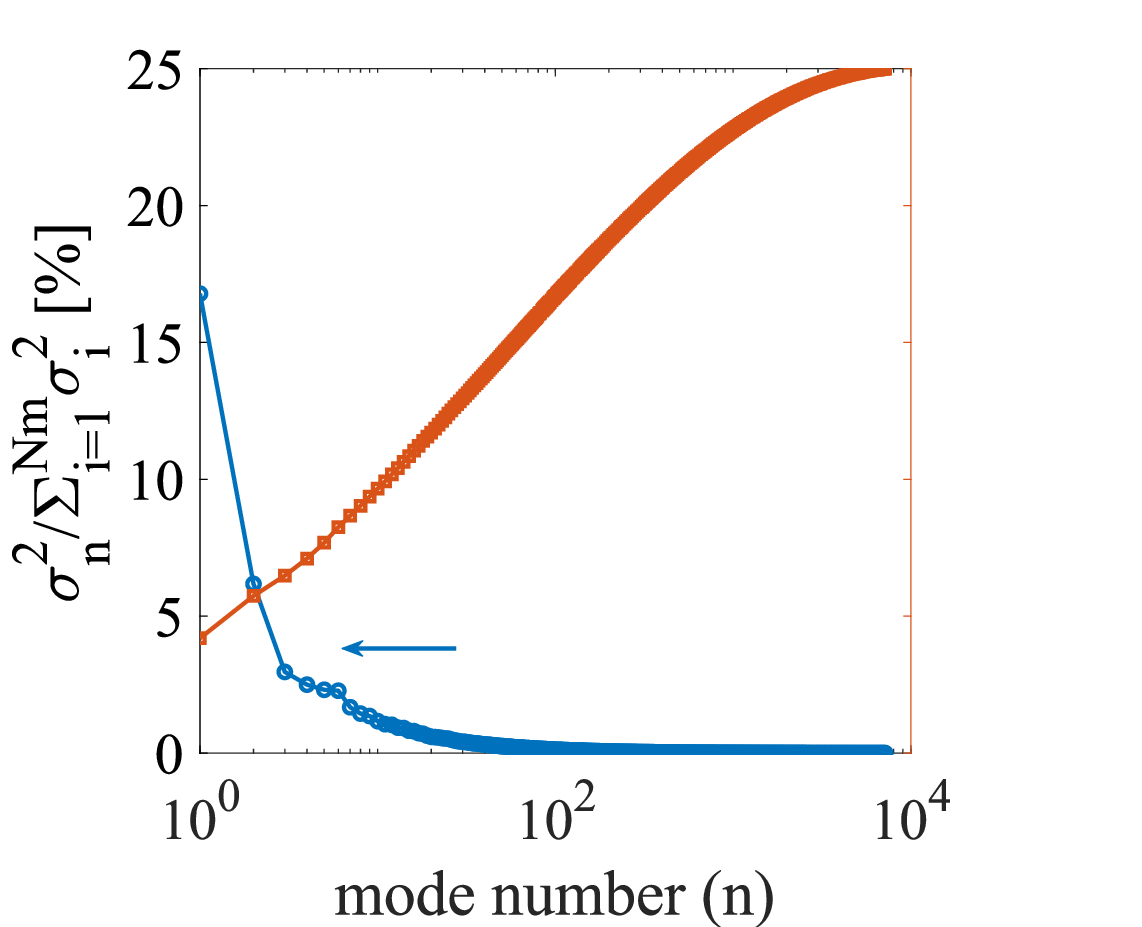}
\end{subfigure}
\hspace{-2mm}
\begin{subfigure}[]
    \centering
    \includegraphics[width=0.48\linewidth,trim=28mm 0mm 2mm 0mm,clip]{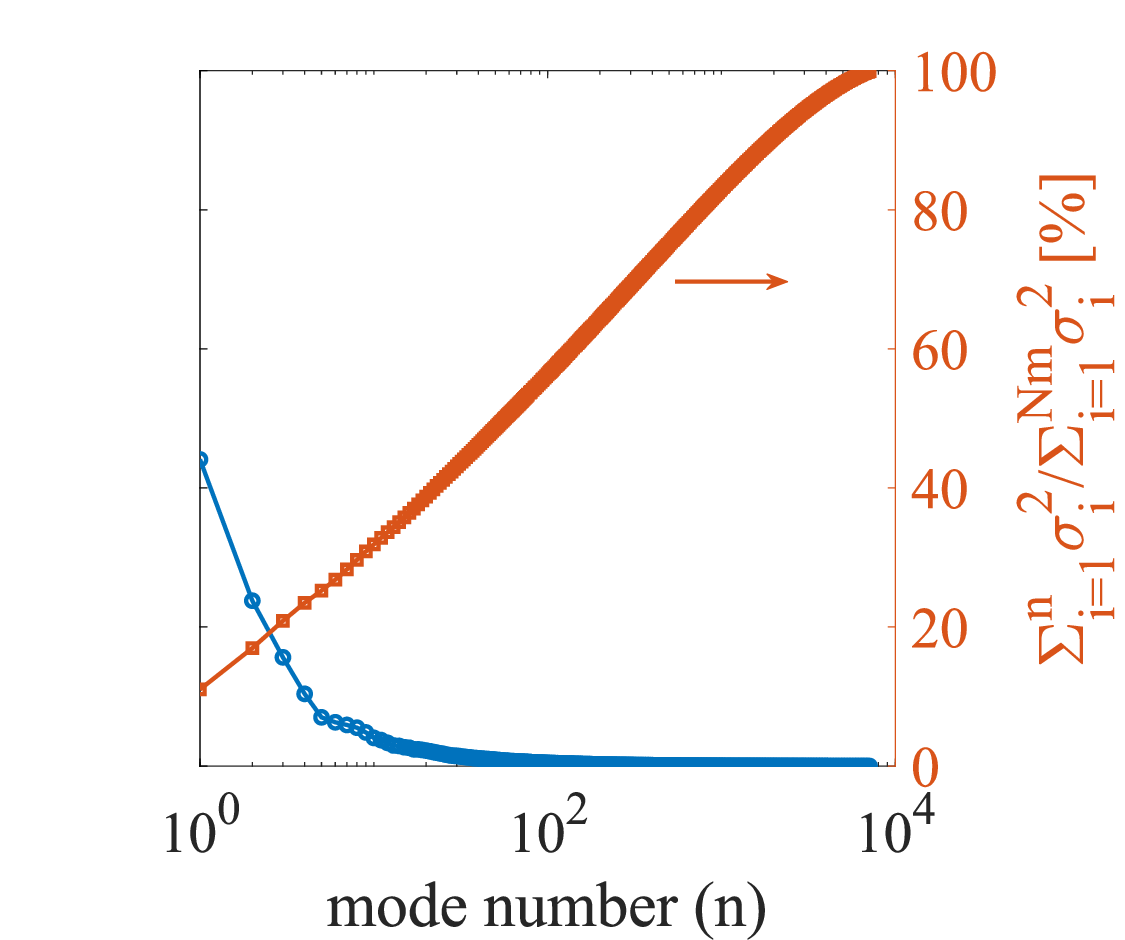}
\end{subfigure}
\caption{The \acs{POD} spectrum from the channel dataset with the sub-domain from (a) \ac{2C} planar velocity field, (b) 3C planar velocity field, (c) 3D velocity field at aspect ratio $1/8$, and (d) 3D velocity field at aspect ratio $1/2$.}
\label{fig: channel_spectrum}
\end{figure}

The \ac{POD} spectrum for the channel flow case is illustrated in Fig. \ref{fig: channel_spectrum}. Compared to the pinball case, the channel spectrum is significantly more dispersed, requiring nearly $1000$ modes to capture $90\%$ of the total energy, whereas the pinball case only needs less than $10$ modes. The estimation of the temporal modes with \ac{EPOD} is thus more severely challenged by the higher dimensionality of the flow. Additionally, for the cases corresponding to thick volumetric domain, the \ac{POD} spectrum spreads out on an even larger number of modes.

\begin{figure}
\centering
\hspace{10mm}
\begin{subfigure}
    \centering
    \includegraphics[width=0.3\linewidth]{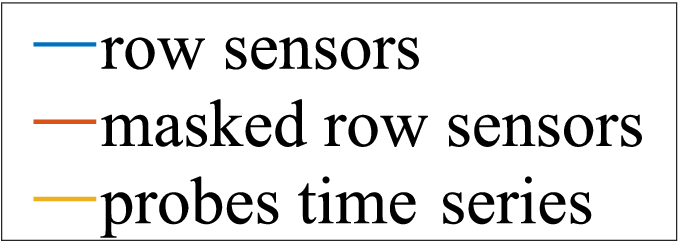}
\end{subfigure}
\newline
\setcounter{subfigure}{0}
\begin{subfigure}[]
    \centering
    \includegraphics[width=0.326\linewidth,trim=8mm 20mm 22mm 4mm,clip]{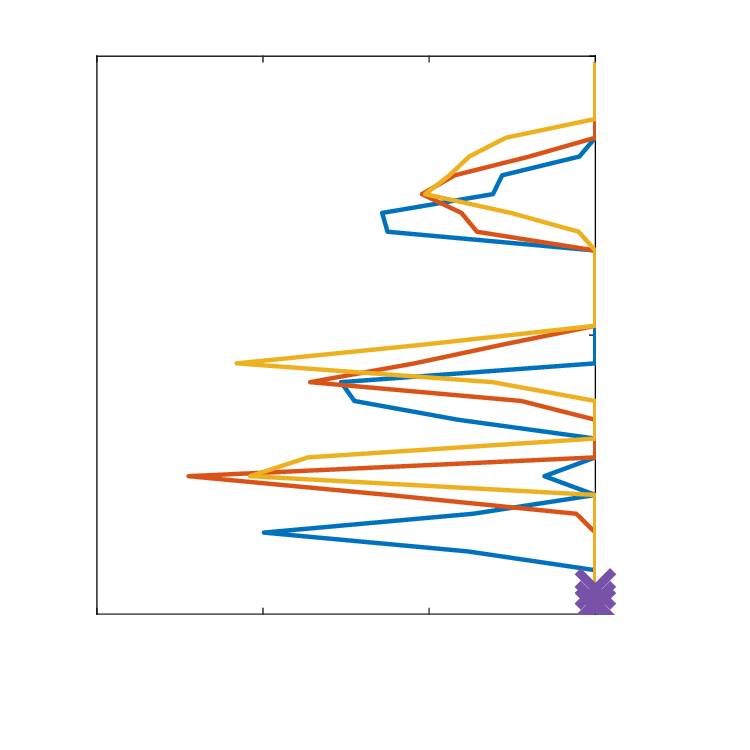}
\end{subfigure}
\hspace{4mm}
\begin{subfigure}[]
    \centering
    \includegraphics[width=0.4\linewidth,trim=8mm 20mm 0mm 4mm,clip]{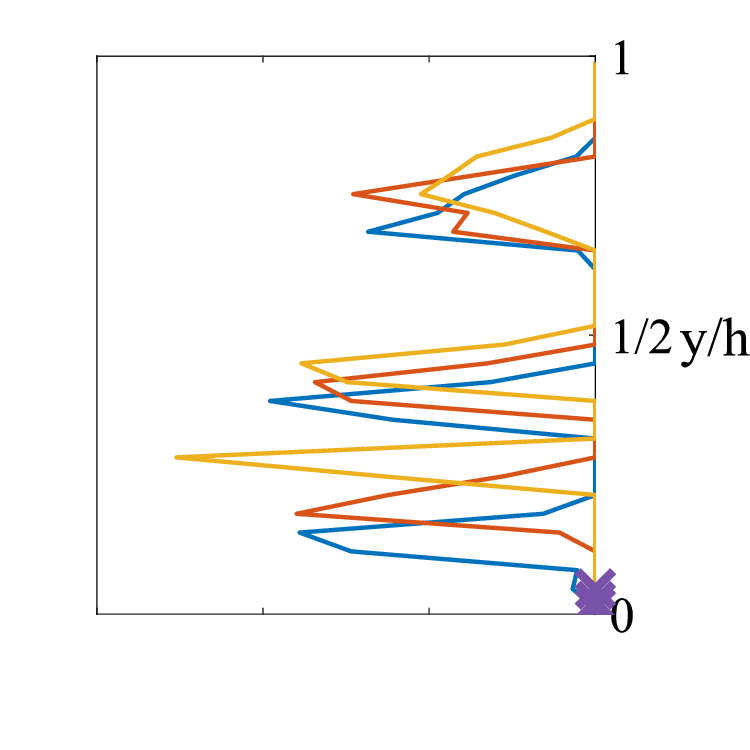}
\end{subfigure}
\begin{subfigure}[]
    \centering
    \includegraphics[width=0.326\linewidth,trim=8mm 0mm 22mm 4mm,clip]{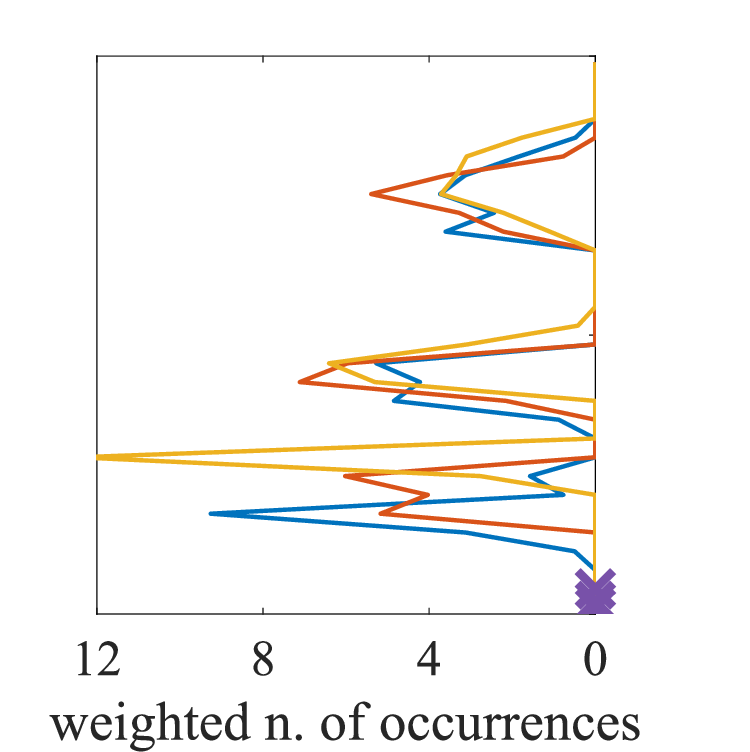}
\end{subfigure}
\hspace{4mm}
\begin{subfigure}[]
    \centering
    \includegraphics[width=0.4\linewidth,trim=8mm 0mm 0mm 4mm,clip]{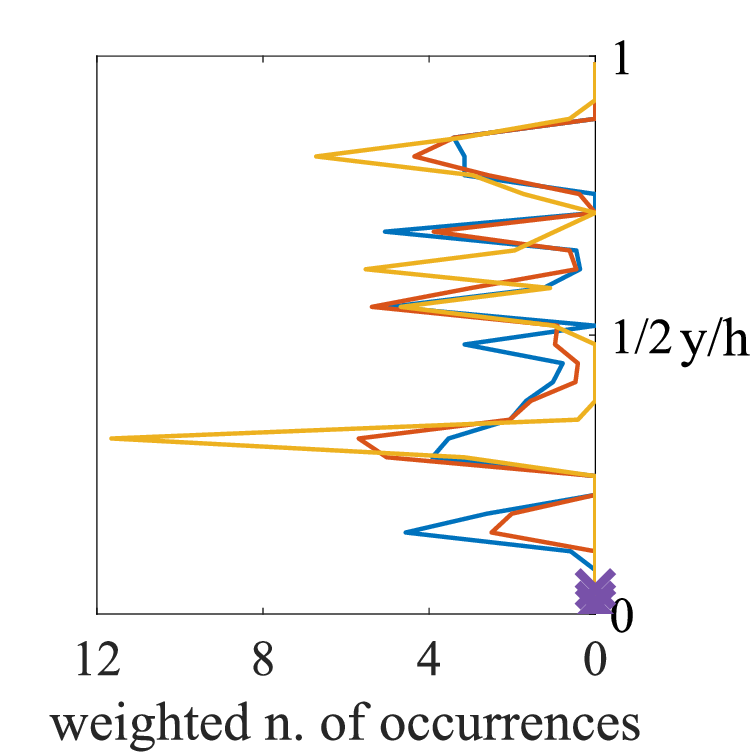}
\end{subfigure}
\caption{The one-dimensional weighted positional frequency distribution of optimal sensor positioning for the channel flow across different aspect ratios of the domain, subfigures show the result from (a) \ac{2C} planar velocity field, (b) 3C planar velocity field, (c) 3D velocity field at aspect ratio $1/8$, and (d) 3D velocity field at aspect ratio $1/2$. The purple $\times$ signs stand for the positioning from block-pivoted QR}
\label{fig: channel_hmap}
\end{figure}

The one-dimensional positional frequency distributions of optimal sensor positioning are shown in Fig. \ref{fig: channel_hmap}. The distribution for all aspect ratios presents $3$ peaks for all types of sensors. We observe that the most performing positions for row surrogate sensors include regions located closer to the wall. This might be ascribed to the higher energy corresponding to that region. The peak positions of probes time series indicate instead that estimation with time-delay embedding is more effective for slightly larger wall-distance. This is likely due to the stronger flow distortion and 3D motion in the near-wall region, which challenge the process of time-delay embedding.  

Once the correlation value is taken into account by masking low correlation region, the optimization with row surrogate sensor provides new locations which are closer to those of the probes time series, and thus arguably with higher potential to be more effective in the process of reconstructing once the probes are placed.

For the 3D-3C test cases, the weighted positional frequency distributions become more irregular since the prediction becomes more complex. Nonetheless, the optimal placements from row sensors and masked row sensors still have similar performance to that from probes time series as listed in Table \ref{tab: channel_error}. It is also worthwhile stressing that the optimal positioning from the planar field also works for the volumetric field with little degradation, which significantly reduces the computational costs of the search.

\begin{figure}[htbp]
\centering
\hspace{20mm}
\begin{subfigure}
    \centering
    \includegraphics[width=0.45\linewidth,trim=10mm 100mm 10mm 16mm,clip]{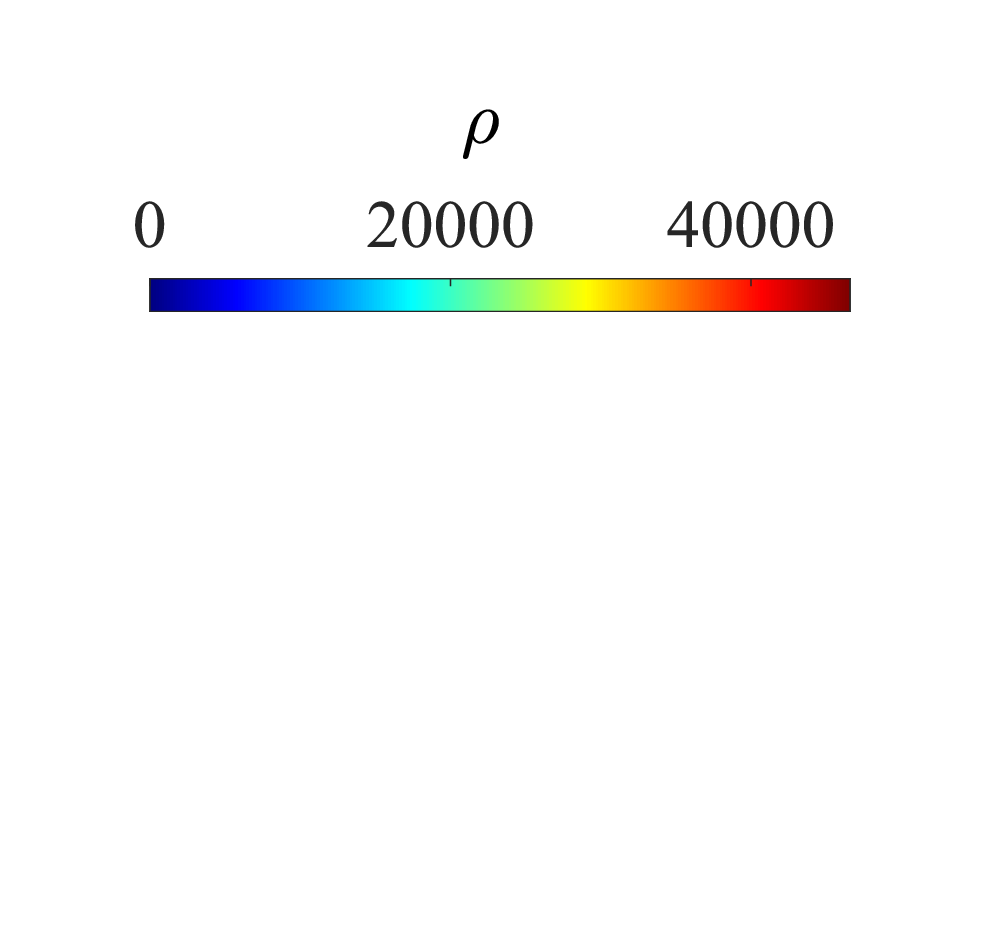}
\end{subfigure}
\newline
\setcounter{subfigure}{0}
\hspace{0mm}
\begin{subfigure}
    \centering
    \includegraphics[width=0.45\linewidth,trim=0mm 24mm 10mm 0mm,clip]{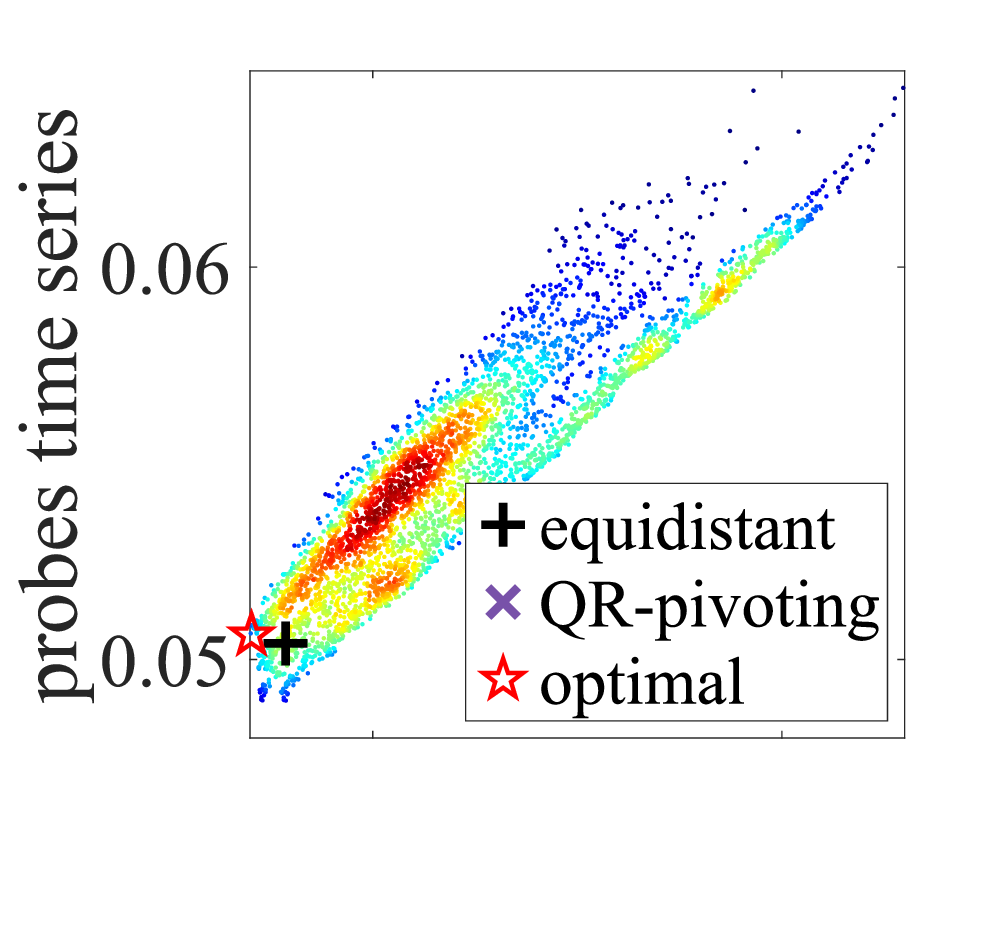}
\end{subfigure}
\hspace{-4mm}
\begin{subfigure}
    \centering
    \includegraphics[width=0.44\linewidth,trim=0mm 24mm 10mm 0mm,clip]{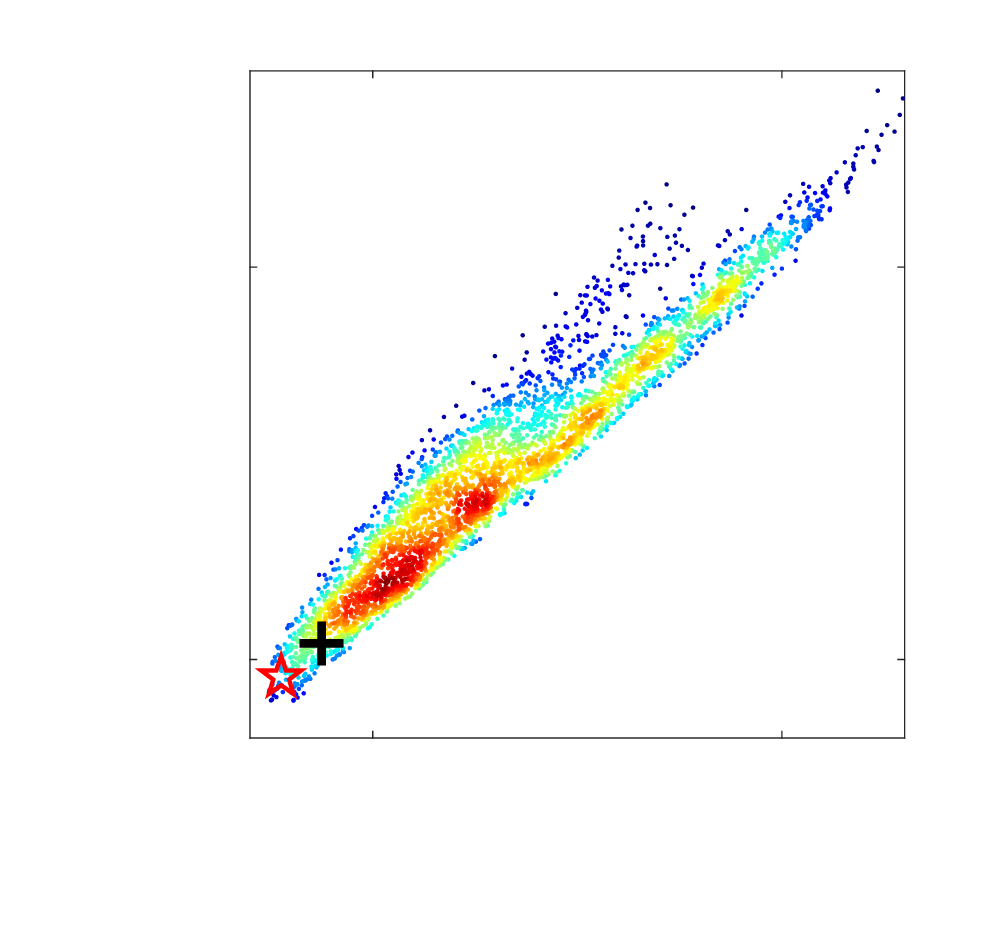}
\end{subfigure}
\begin{subfigure}
    \centering
    \includegraphics[width=0.45\linewidth,trim=0mm 24mm 10mm 0mm,clip]{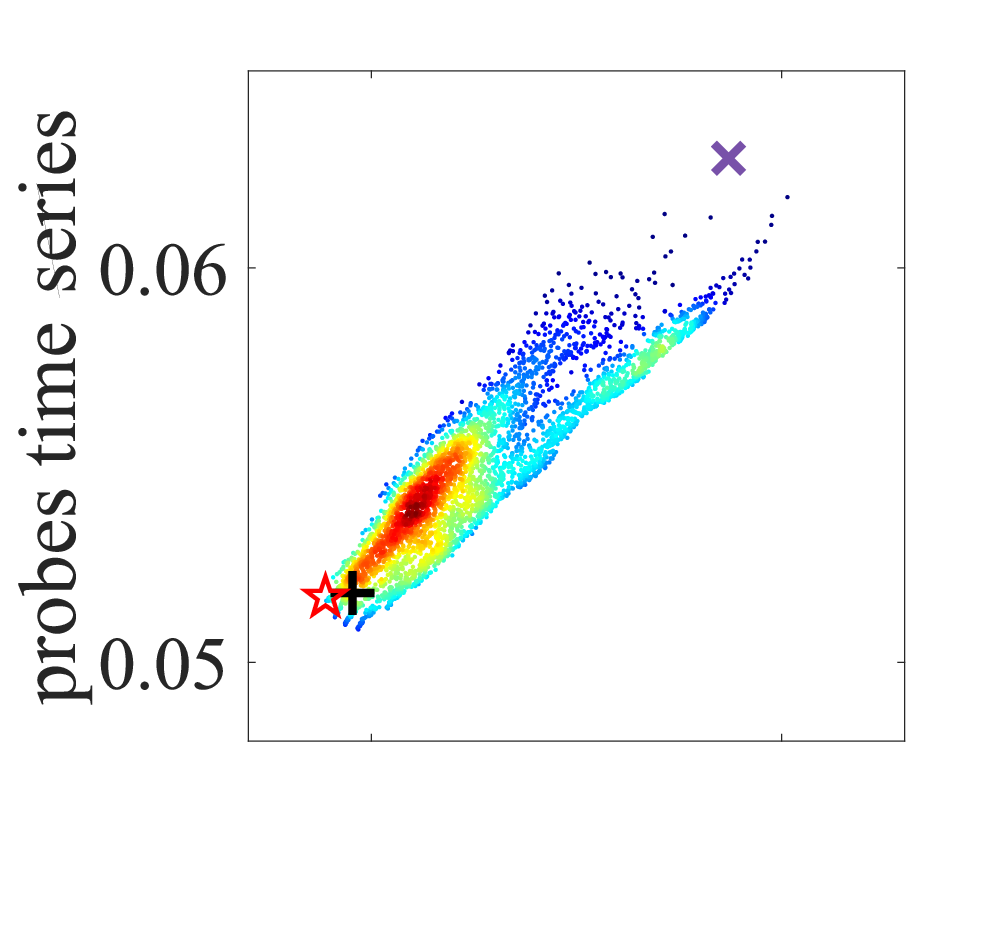}
\end{subfigure}
\hspace{-4mm}
\begin{subfigure}
    \centering
    \includegraphics[width=0.45\linewidth,trim=0mm 24mm 10mm 0mm,clip]{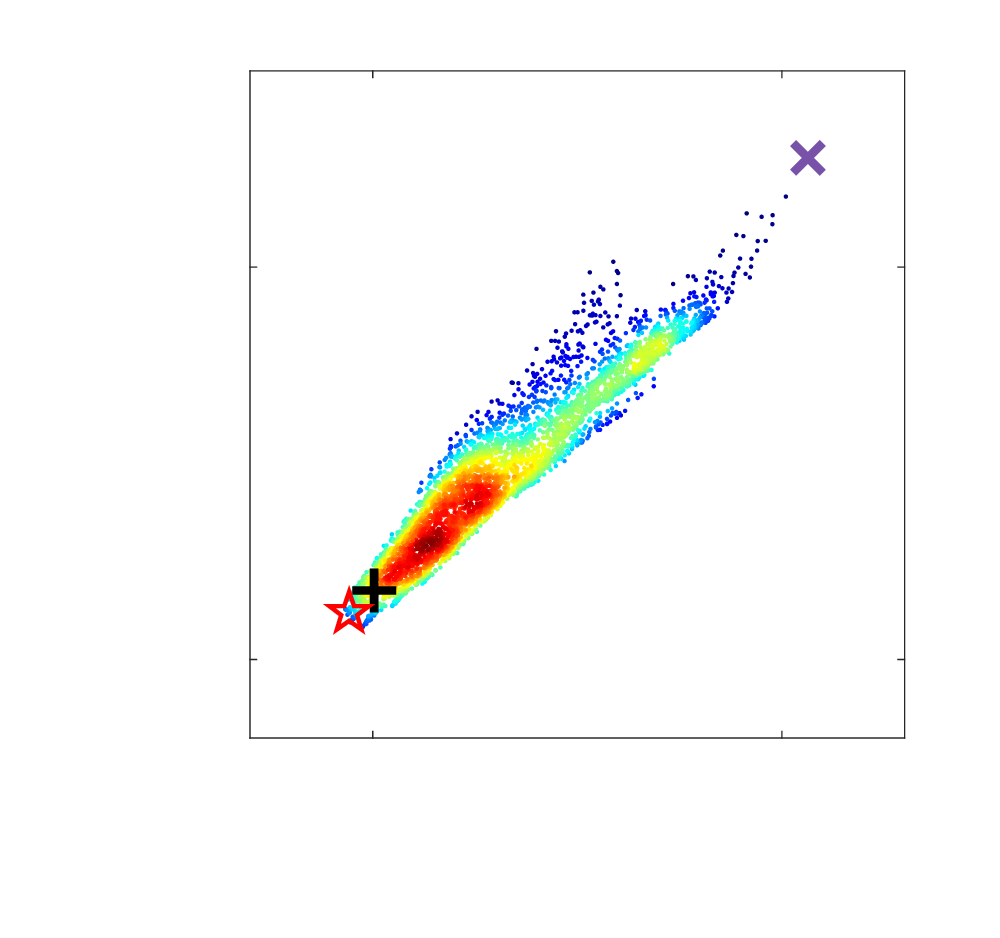}
\end{subfigure}
\begin{subfigure}
    \centering
    \includegraphics[width=0.45\linewidth,trim=0mm 24mm 10mm 0mm,clip]{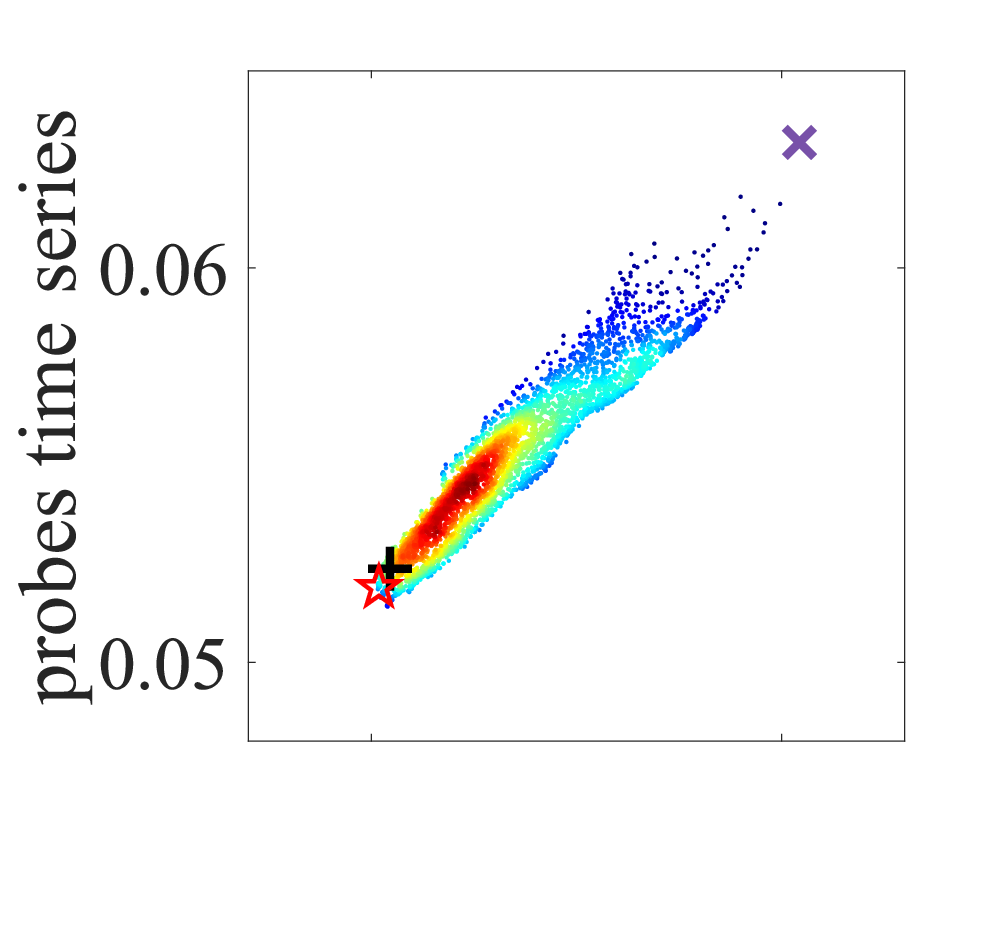}
\end{subfigure}
\hspace{-4mm}
\begin{subfigure}
    \centering
    \includegraphics[width=0.45\linewidth,trim=0mm 24mm 10mm 0mm,clip]{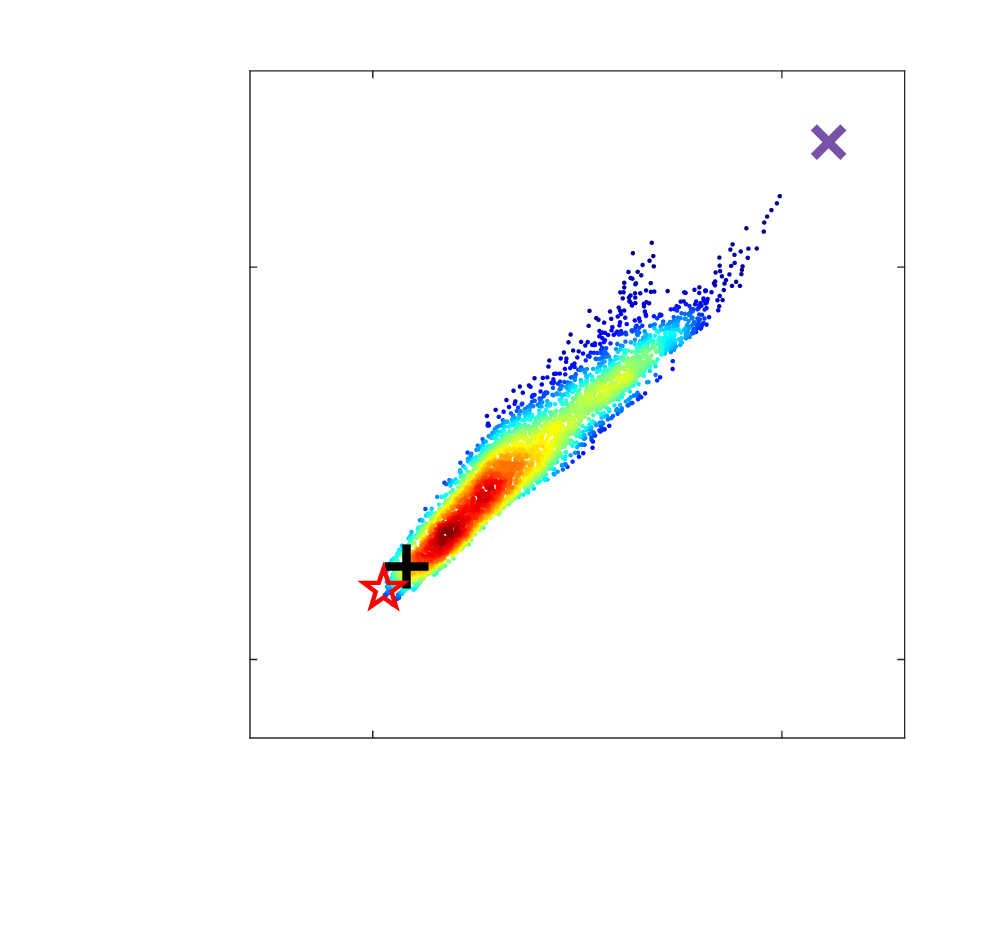}
\end{subfigure}
\begin{subfigure}
    \centering
    \includegraphics[width=0.45\linewidth,trim=0mm 0mm 10mm 0mm,clip]{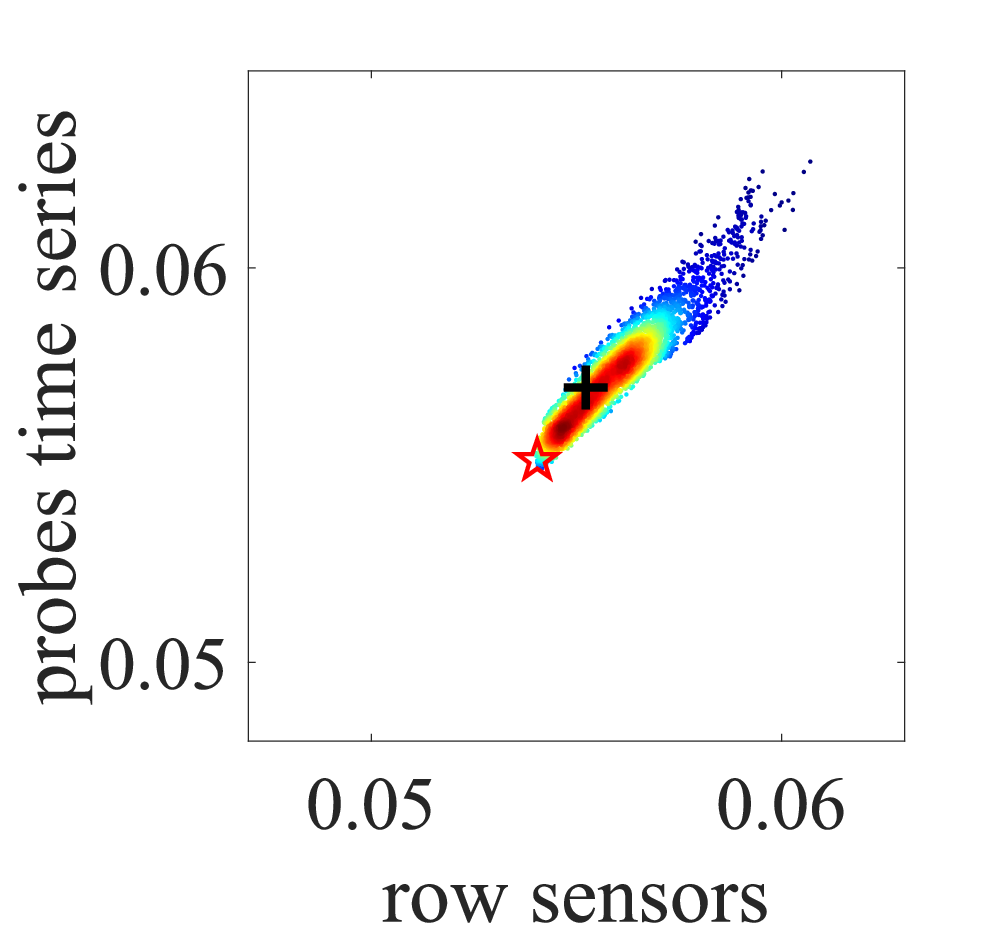}
\end{subfigure}
\hspace{-4mm}
\begin{subfigure}
    \centering
    \includegraphics[width=0.45\linewidth,trim=0mm 0mm 10mm 0mm,clip]{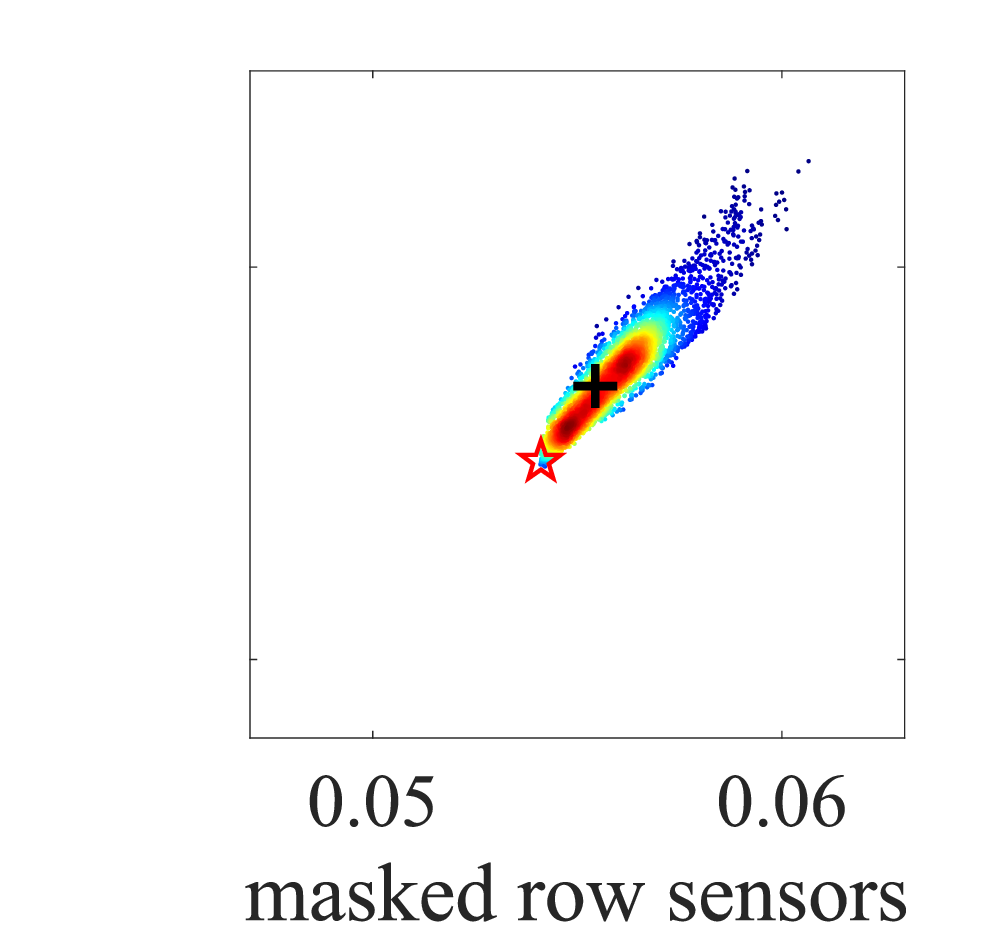}
\end{subfigure}
\caption{The 2D probability density function of the velocity reconstruction error using probes with time-series versus row sensors (left) or row sensors masked (right). The computation domains are, from top to bottom: \ac{2C} planar (1st row), 3C planar (2nd row), thin volumetric with aspect ratio $1/8$ (3rd row), volumetric with aspect ratio $1/2$ (4th row). The colour of the spots represents local density ranging from low (blue) to high (red). It is set $\displaystyle\sum_i\rho\Delta s_i = 1$, where the $\Delta s_i$ is any local area on the scatter plots.}
\label{fig: channel_scatter}
\end{figure}

The scatter plots in Fig. \ref{fig: channel_scatter} reflect a high similarity in velocity reconstruction between using probes time series and row surrogate sensors. The process of masking further improves this correlation, with a sharper downleft corner in almost all cases. The optimal positioning given by masked row sensors is very close to that of probes time series. This indicates that row sensors after masking are good surrogates for the probes time series. 

The equidistant probes provide also reasonably good positioning for all tested cases, even slightly outperforming the row surrogate sensors positioning in planar fields. The block-pivoted QR, on the other hand, provides poorer performances, possibly because it focuses on placing probes at sites with the most information as shown in Fig. \ref{fig: channel_hmap} (a) (i.e. region very close to the wall), rather than covering a broader area.

\begin{table*}[]
\centering
\begin{tabu}{|[2pt]l|[2pt]c c c c|[2pt]c|[2pt]}
    \tabucline[2pt]{-}
    \multirow{2}{*}{\diagbox{field type}{probe type}} & \bf{equidistant} & \bf{block-} & \bf{surrogate} & \bf{surrogate masked} & \bf{probes} \\ 
    & \bf{probes} & \bf{pivoted QR} & \bf{row sensors} & \bf{row sensors} & \bf{time series} \\
    \tabucline[2pt]{-}
    \bf{planar PIV}  & 0.0504 & 0.0660 & 0.0506 & 0.0496 & 0.0490 \\
    \hline
    \bf{stereoscopic PIV} & 0.0518 & 0.0628 & 0.0516 & 0.0512 & 0.0508 \\
    \hline
    \bf{thin tomographic PIV:} &  &  &  &  &  \\
    with 3D positioning & 0.0524 & 0.0632 & 0.0519 & 0.0518 & 0.0514 \\
    with 2D positioning & & & 0.0522 & 0.0516 & \\
    \hline
    \bf{thick tomographic PIV:} &  &  &  &  &  \\
    with 3D positioning & 0.0570 & 0.0669 & 0.0551 & 0.0550 & 0.0549 \\
    with 2D positioning & & & 0.0553 & 0.0554 & \\
    \tabucline[2pt]{-}
\end{tabu}
\caption{The error of velocity reconstruction using time-delay embedding and sensors located according to different methods for the turbulent channel flow. For the volumetric test case, we consider either exploring the entire 2D downstream end of the domain (in the table as ``3D positioning'') or solely the central slice (``2D positioning''). The error is defined as in Eq. \ref{eq: error}. In the last column, the optimal case of online optimization with probes time series is included for reference.}
\label{tab: channel_error}
\end{table*}

\subsection{3D experiment of propeller wake}

The proposed method is finally tested on the velocity field from time-resolved tomographic PIV measurements of the wake of a rotating propeller. To have a solid ground truth for comparison, the original sequence is subsampled in time to produce a non-time-resolved sequence. Probes with time series are then generated at any node at the downstream edge of the domain by using data from the original time-resolved sequence. For the analysis, we consider planar data obtained by extracting the central slice of the volume, and the full volumetric measurements.

\begin{figure*}
\centering
\begin{subfigure}
    \centering
    \raisebox{-7mm}{
    \includegraphics[width=0.5\linewidth,trim=4mm 0mm 32mm 0mm,clip]{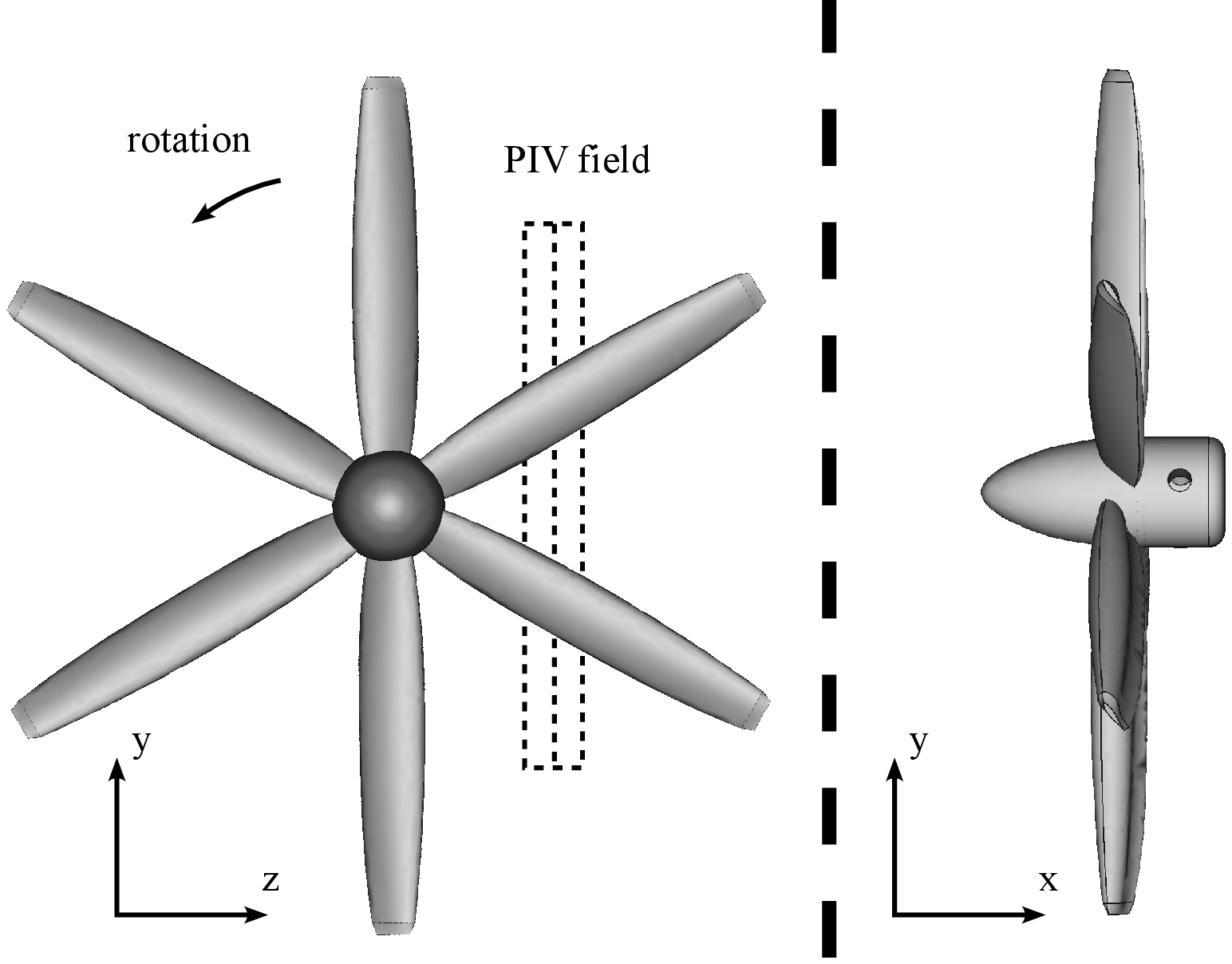}}
\end{subfigure}
\hspace{0mm}
\setcounter{subfigure}{0}
\begin{subfigure}[]
    \centering
    \includegraphics[width=0.2\linewidth,trim=0mm 0mm 32mm 0mm,clip]{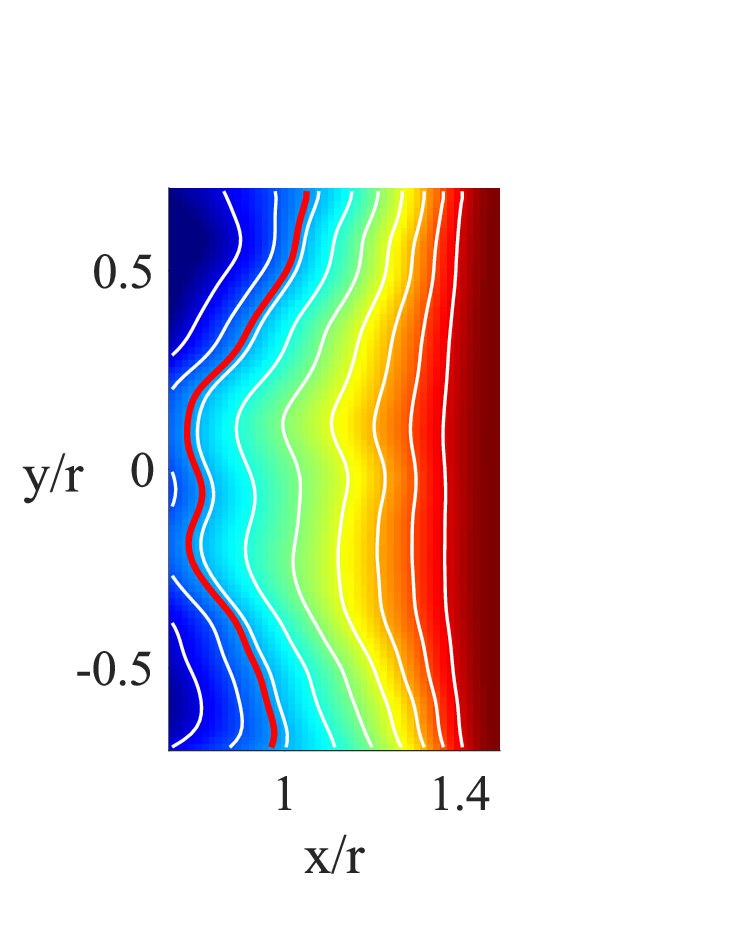}
\end{subfigure}
\begin{subfigure}[]
    \centering
    \includegraphics[width=0.2\linewidth,trim=32mm 0mm 0mm 0mm,clip]{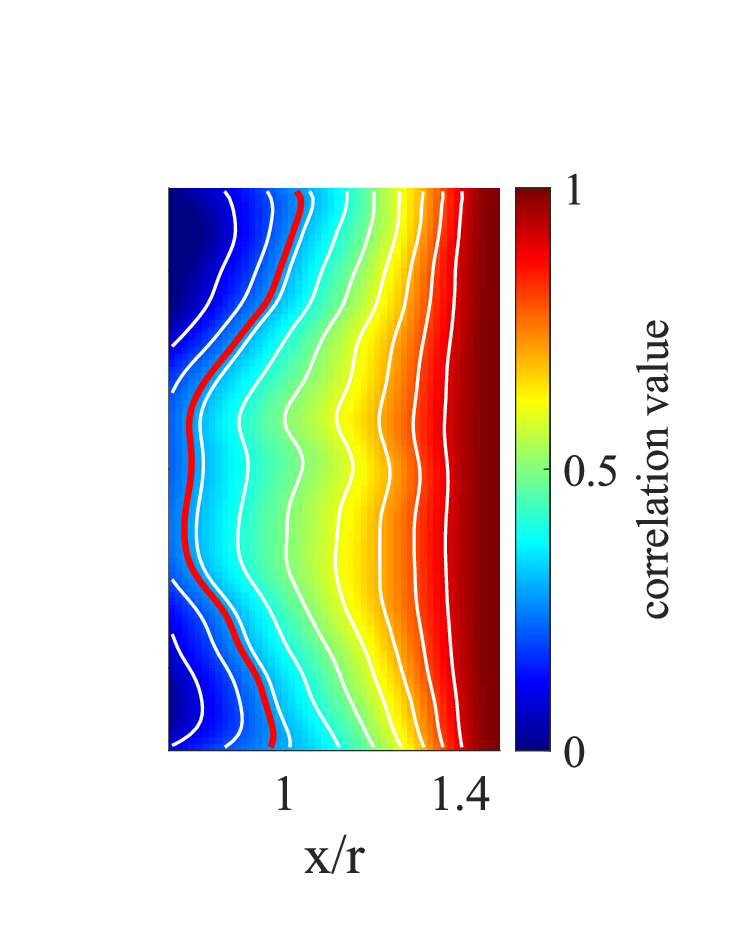}
\end{subfigure}
\caption{The front and left view of the blades model and the correlation map for (a) planar \ac{PIV} obtained by extracting data from the centre slice of the tomographic \ac{PIV} domain and (b) by averaging along the depth direction the correlation map from the full tomographic \ac{PIV} field. The thick curves shows the left margin of the region after masking by correlation.}
\label{fig: blades_cmap}
\end{figure*}

\begin{figure}
\centering
\begin{subfigure}[]
    \centering
    \includegraphics[width=0.7\linewidth,trim=0mm 16mm 0mm 0mm,clip]{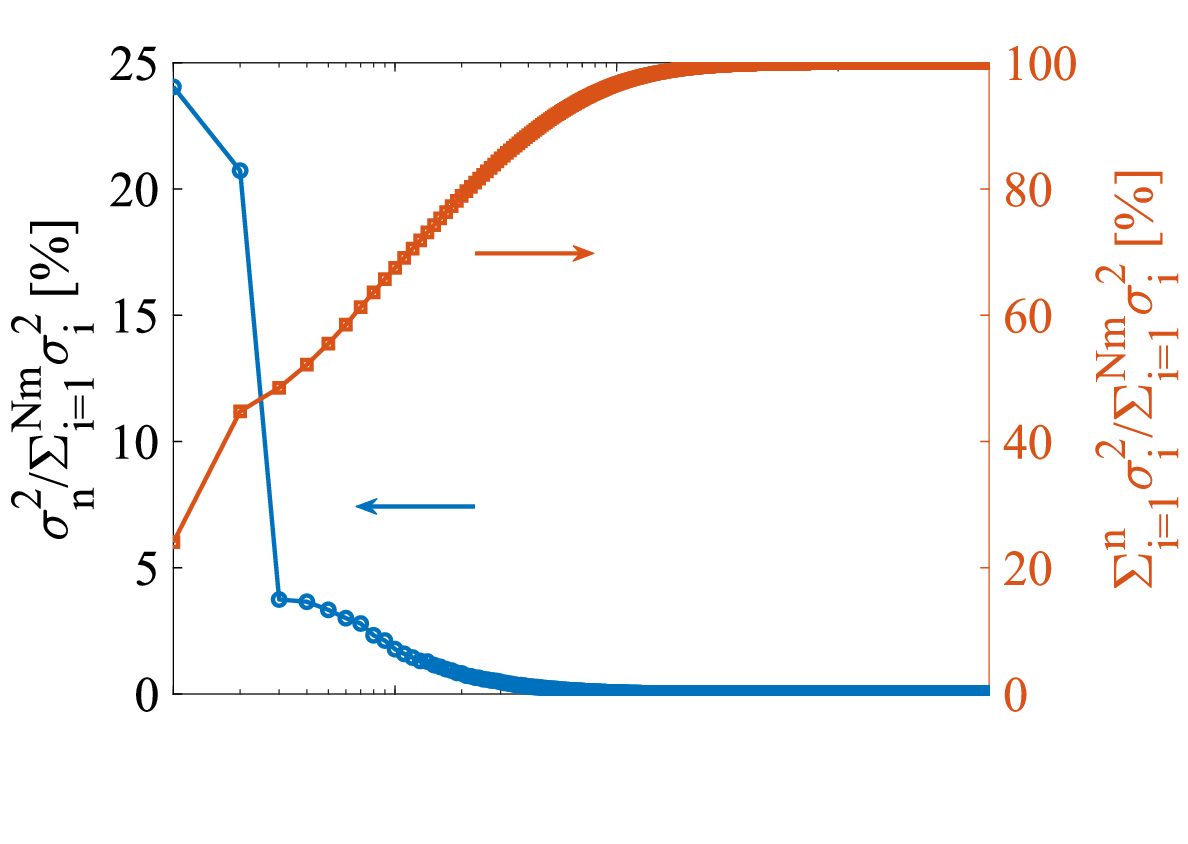}
\end{subfigure}
\begin{subfigure}[]
    \centering
    \includegraphics[width=0.7\linewidth,trim=0mm 16mm 0mm 0mm,clip]{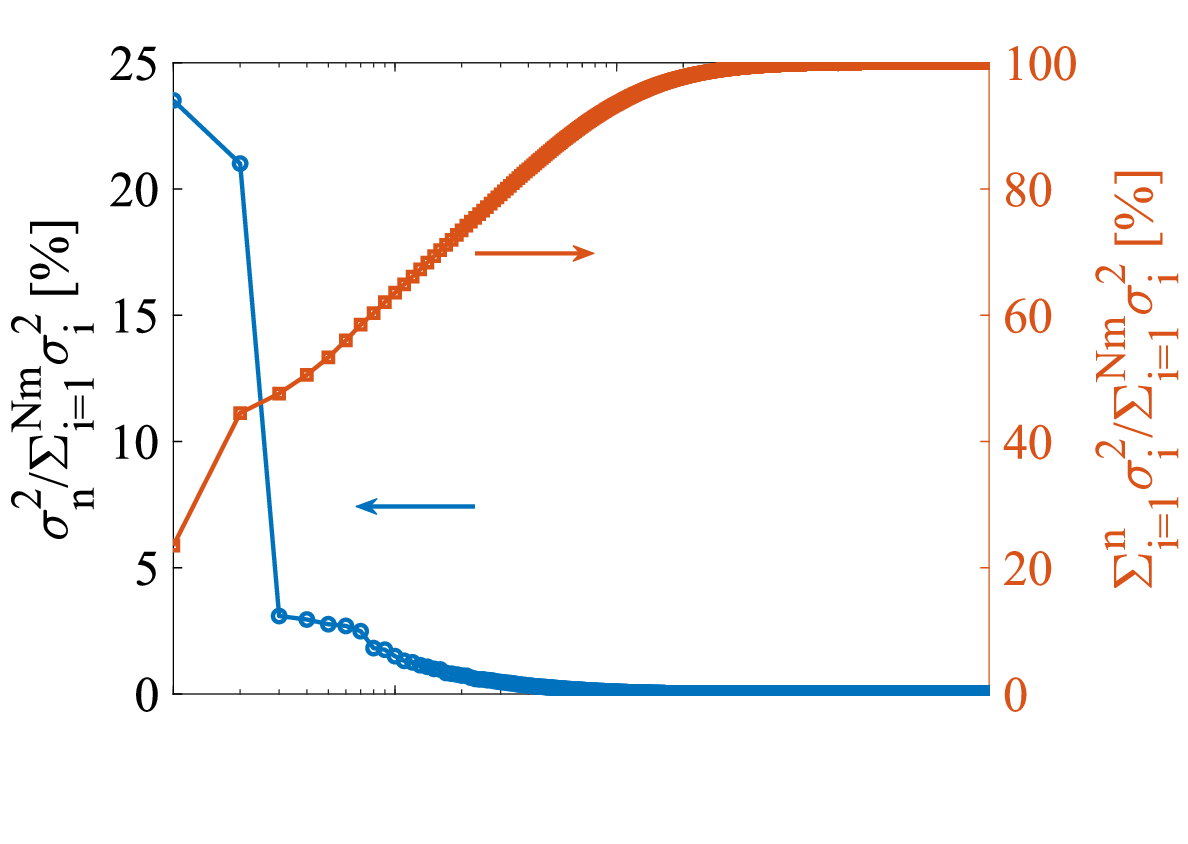}
\end{subfigure}
\begin{subfigure}[]
    \centering
    \includegraphics[width=0.7\linewidth,trim=0mm 0mm 0mm 0mm,clip]{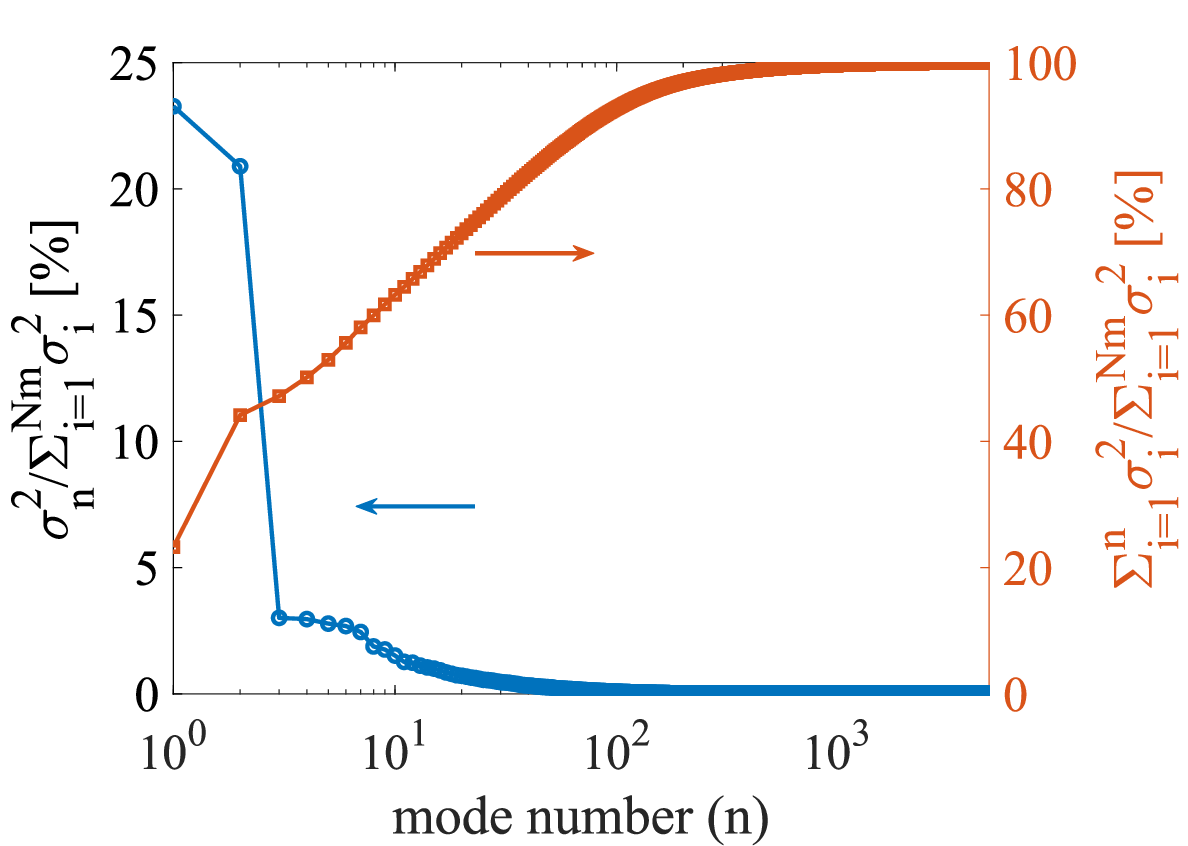}
\end{subfigure}
\caption{The POD spectrum from the blade dataset with the sub-domain from (\textbf{a}) planar velocity field, (\textbf{b}) 3C planar velocity field, (\textbf{c}) thin tomographic field.}
\label{fig: blades_spectrum}
\end{figure}

The correlation maps of the central slice and volumetric field with respect to the points on the downstream edge of the domain have similar distribution, as shown in Fig. \ref{fig: blades_cmap}. In both cases, slightly larger correlation values are observed in the wake of the hub of the windmilling propeller. The \ac{POD} spectra in Fig. \ref{fig: blades_spectrum} reveal a more dispersed distribution if compared to the pinball case, thus indicating a higher level of complexity. The noise in \ac{PIV} measurements further broadens the POD eigenspectrum. No significant differences are observed between planar and volumetric measurements.

\begin{figure}
\centering
\hspace{10mm}
\begin{subfigure}
    \centering
    \includegraphics[width=0.3\linewidth]{fig/legend.eps}
\end{subfigure}
\newline
\hspace{0mm}
\setcounter{subfigure}{0}
\begin{subfigure}[]
    \centering
    \includegraphics[width=0.6\linewidth,trim=12mm 12mm 4mm 8mm,clip]{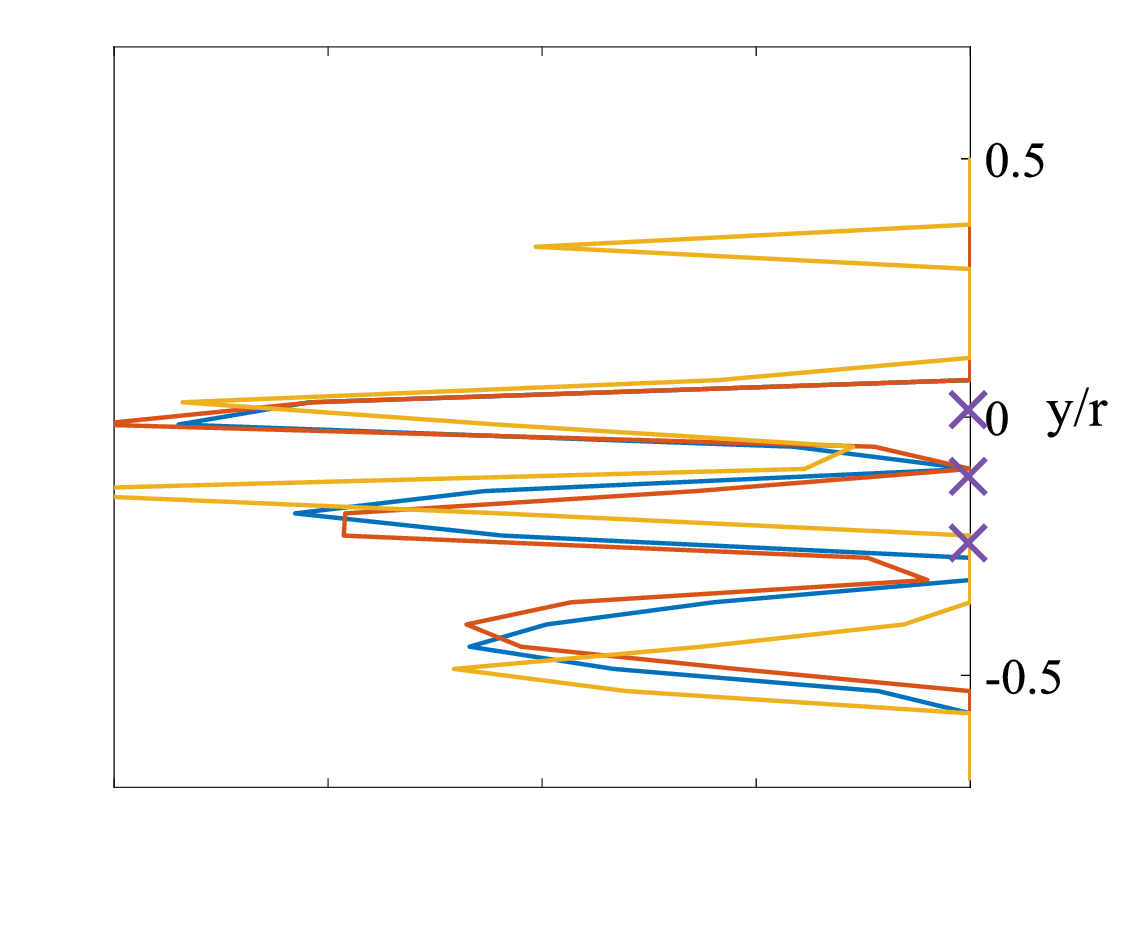}
\end{subfigure}
\hspace{-6mm}
\begin{subfigure}[]
    \centering\includegraphics[width=0.6\linewidth,trim=12mm 12mm 4mm 8mm,clip]{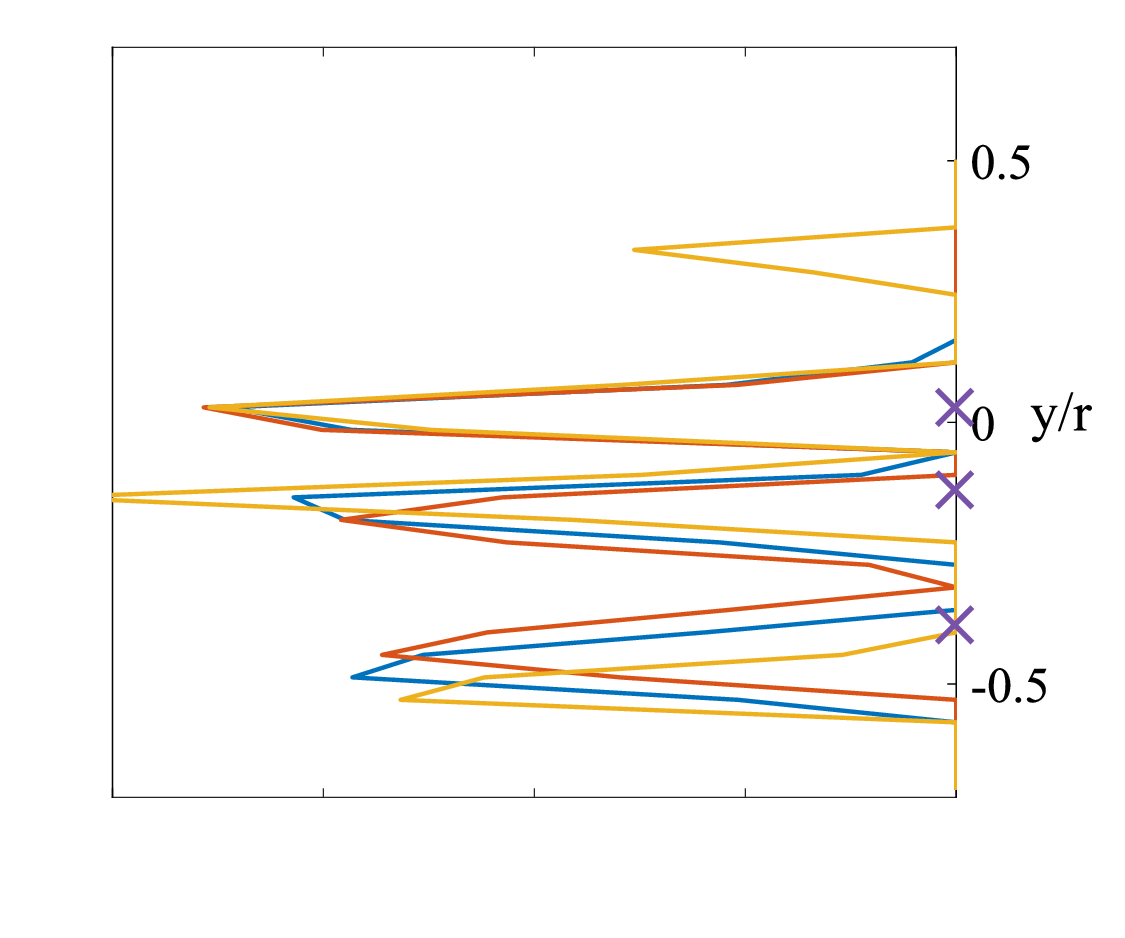}
\end{subfigure}
\hspace{-6mm}
\begin{subfigure}[]
    \centering
    \includegraphics[width=0.6\linewidth,trim=12mm 0mm 4mm 8mm,clip]{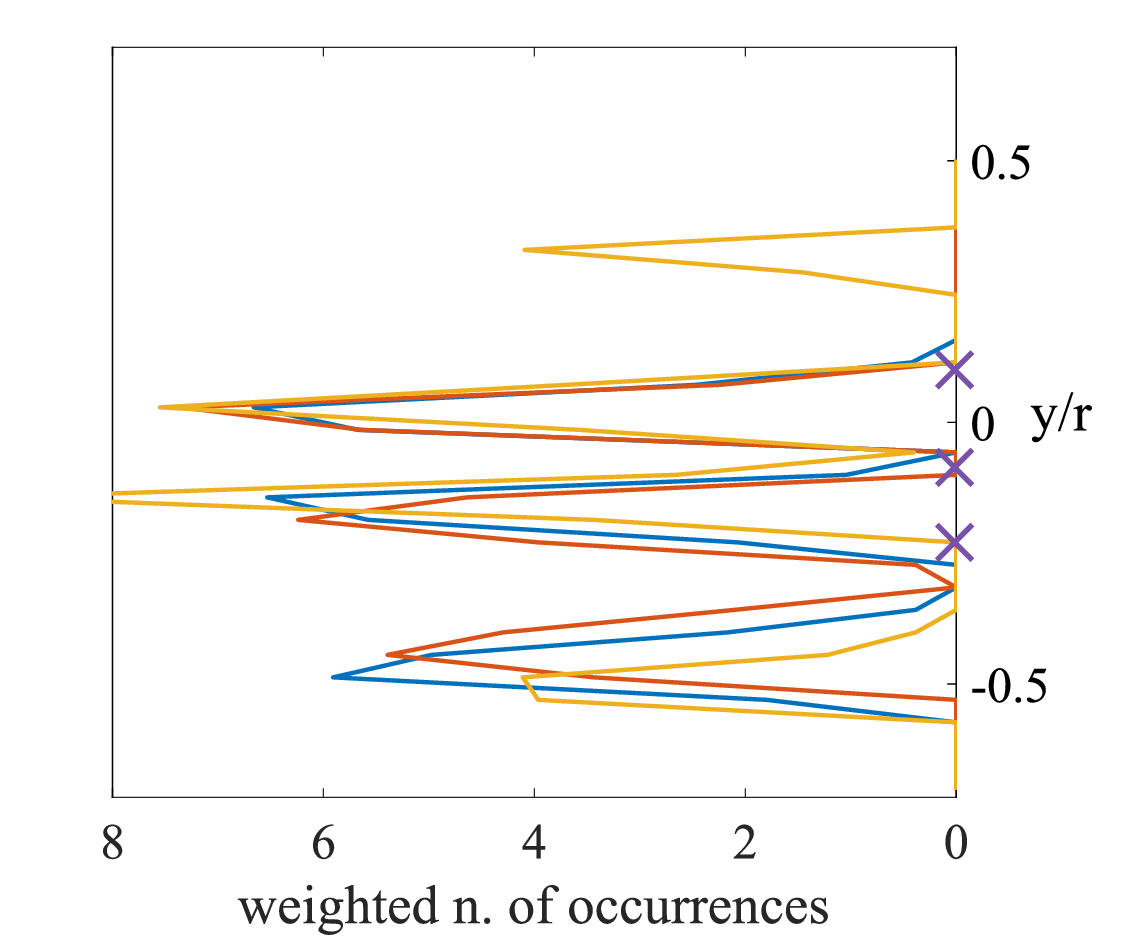}
\end{subfigure}
\caption{The one-dimensional weighted positional distribution of optimal sensor positioning after testing with row sensors, row sensors masked and probes with time-series: (a) \ac{2C} planar \ac{PIV}; (b) 3C planar \ac{PIV}; (c) tomographic \ac{PIV} field. The purple $\times$ signs stand for the positioning from block-pivoted QR.}
\label{fig: blades_hmap}
\end{figure}

Due to the limited field of view in the experiment, the whole domain of this case is immersed in the wake of the windmilling propeller.  As shown in Fig. \ref{fig: blades_hmap}, the most frequent distributions of positions are similar for probes time series and row surrogate sensors (with and without masking). The distribution from probes time series is almost symmetric with respect to the position of the axis of the blades. We observe $4$ peaks in the positional frequency distribution, although the peak located at $y/r = 0.3$ is weaker. On the other hand, both row sensors and masked row sensors locate the optimal sensor position near the main peaks from probes time series. 

The velocity prediction errors using row sensors and masked row sensors in Table \ref{tab: blade_error} are close to the ones from probes time series. As a result, the row sensors, either masked or not, demonstrate to be a good surrogate to the probe time series also in this case. This demonstrates that a preliminary experiment using planar \ac{PIV} is sufficient for sensor positioning in the data-driven flow estimation even if later the experiment involves thin tomographic \ac{PIV} measurements.

\begin{figure}[h]
\centering
\hspace{20mm}
\begin{subfigure}
    \centering
    \includegraphics[width=0.4\linewidth,trim=10mm 100mm 10mm 16mm,clip]{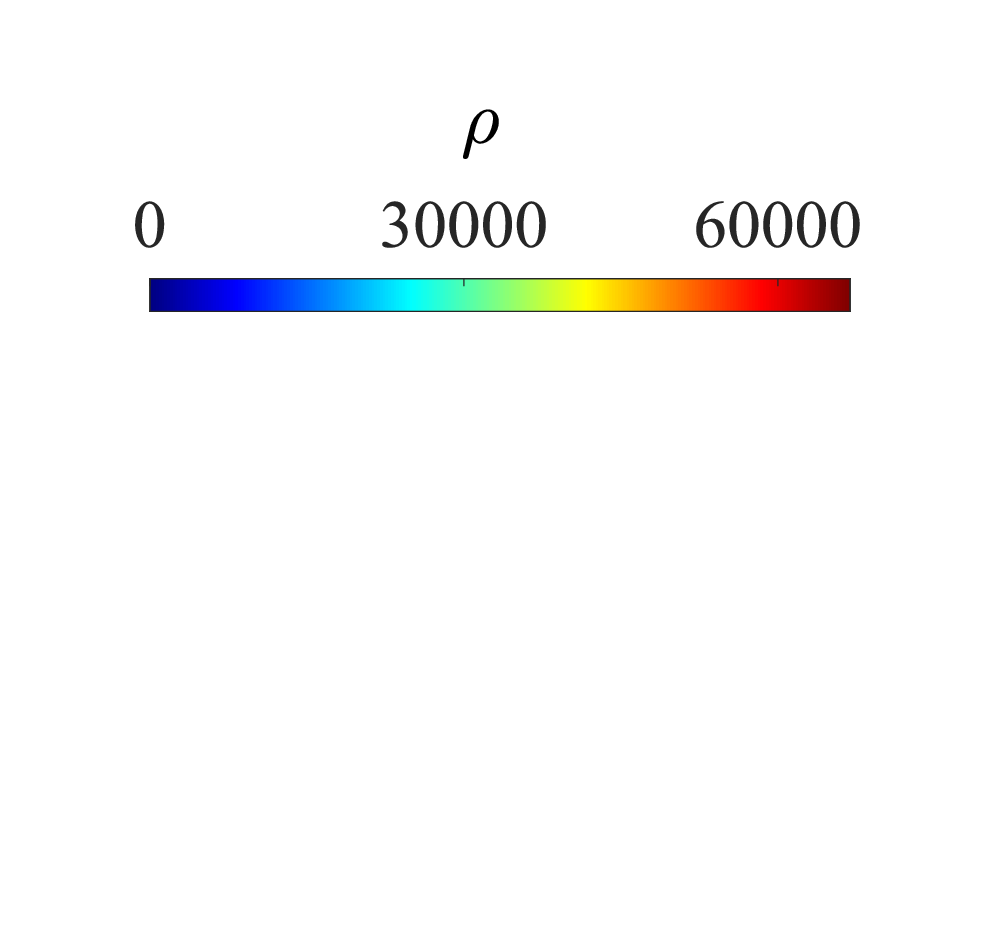}
\end{subfigure}
\newline
\setcounter{subfigure}{0}
\begin{subfigure}
    \centering
    \includegraphics[width=0.42\linewidth,trim=0mm 24mm 16mm 0mm,clip]{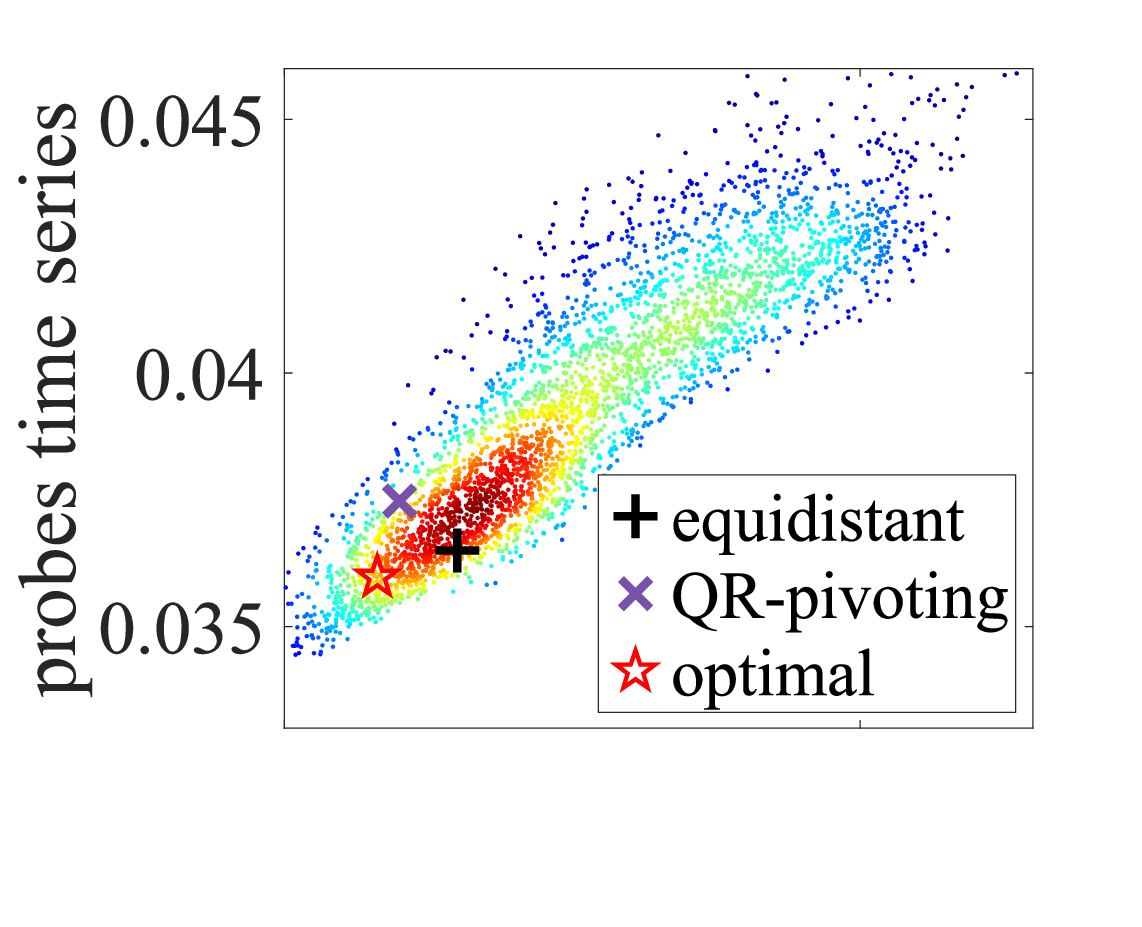}
\end{subfigure}
\begin{subfigure}
    \centering
    \includegraphics[width=0.37\linewidth,trim=24mm 24mm 12mm 0mm,clip]{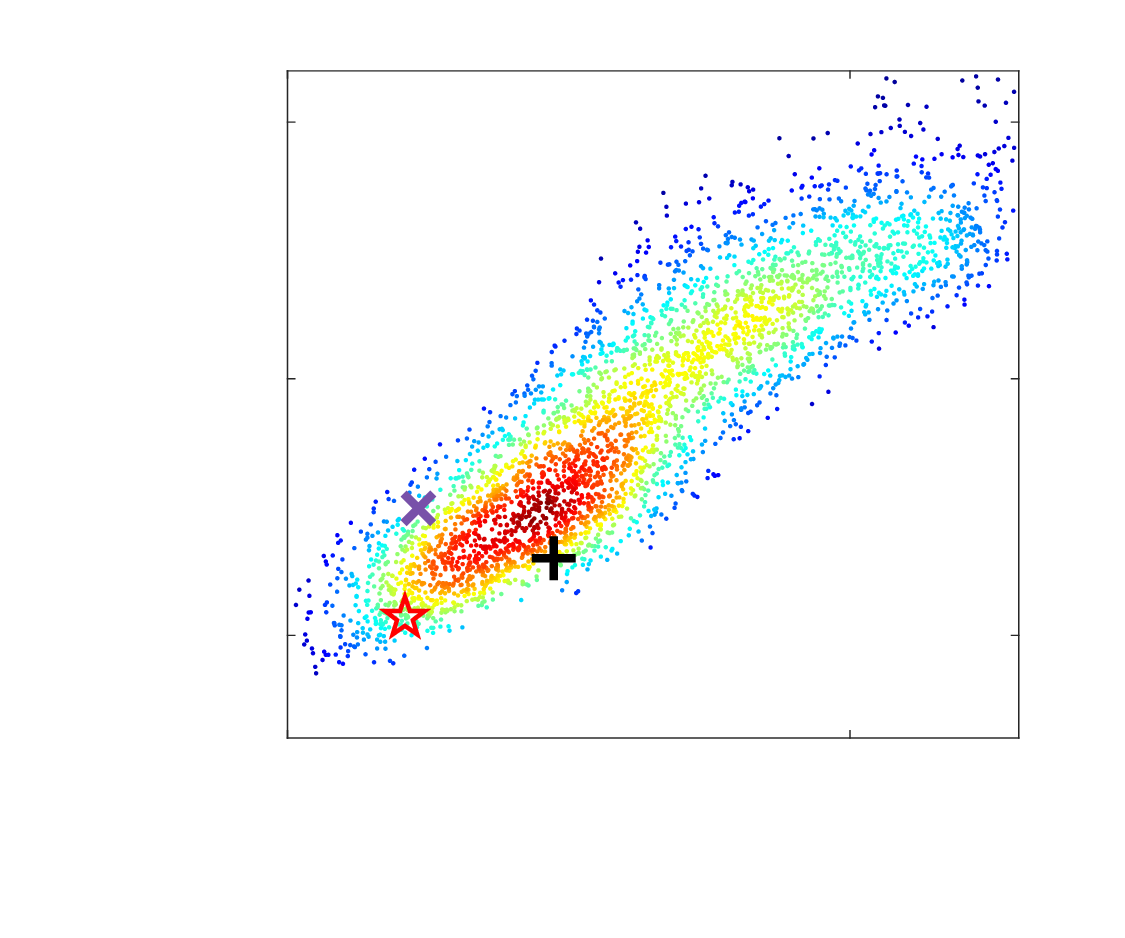}
\end{subfigure}
\begin{subfigure}
    \centering
    \includegraphics[width=0.42\linewidth,trim=0mm 24mm 16mm 0mm,clip]{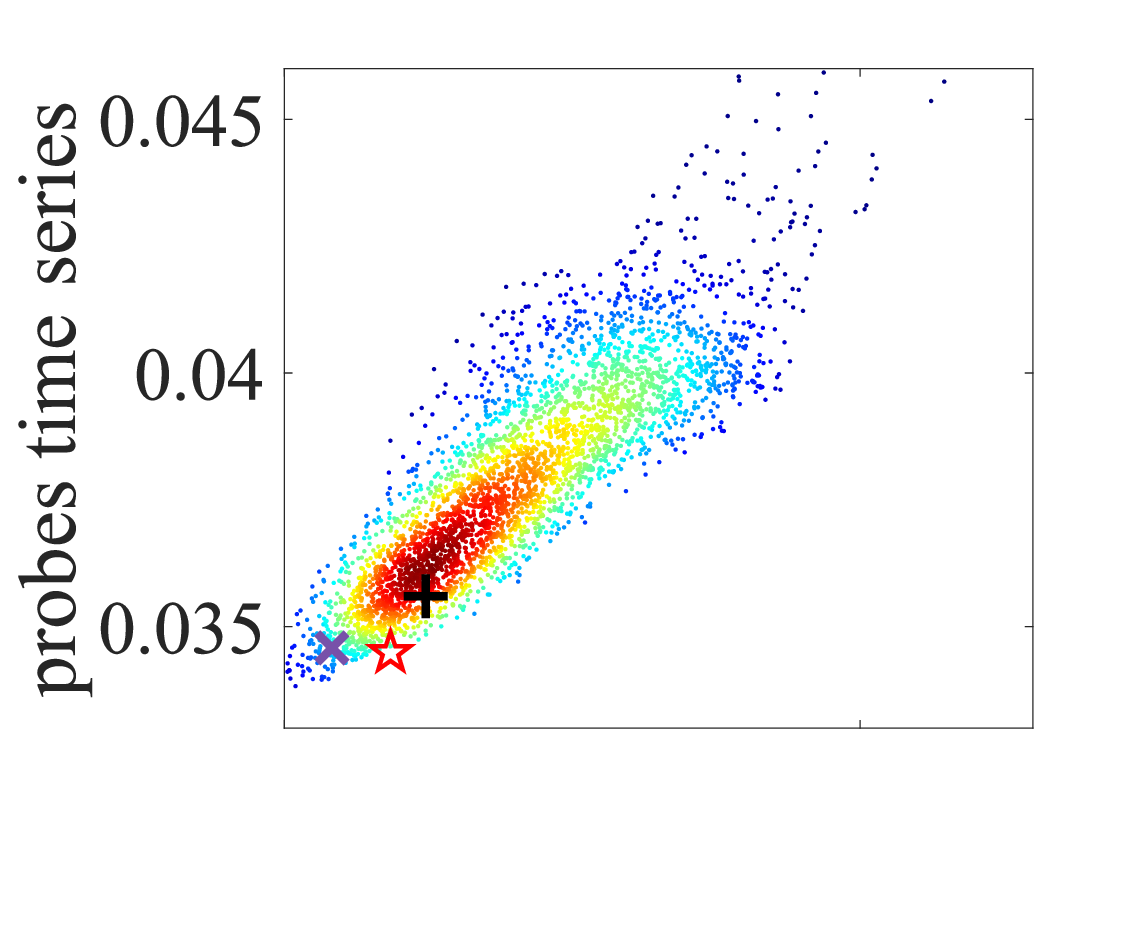}
\end{subfigure}
\begin{subfigure}
    \centering
    \includegraphics[width=0.37\linewidth,trim=24mm 24mm 12mm 0mm,clip]{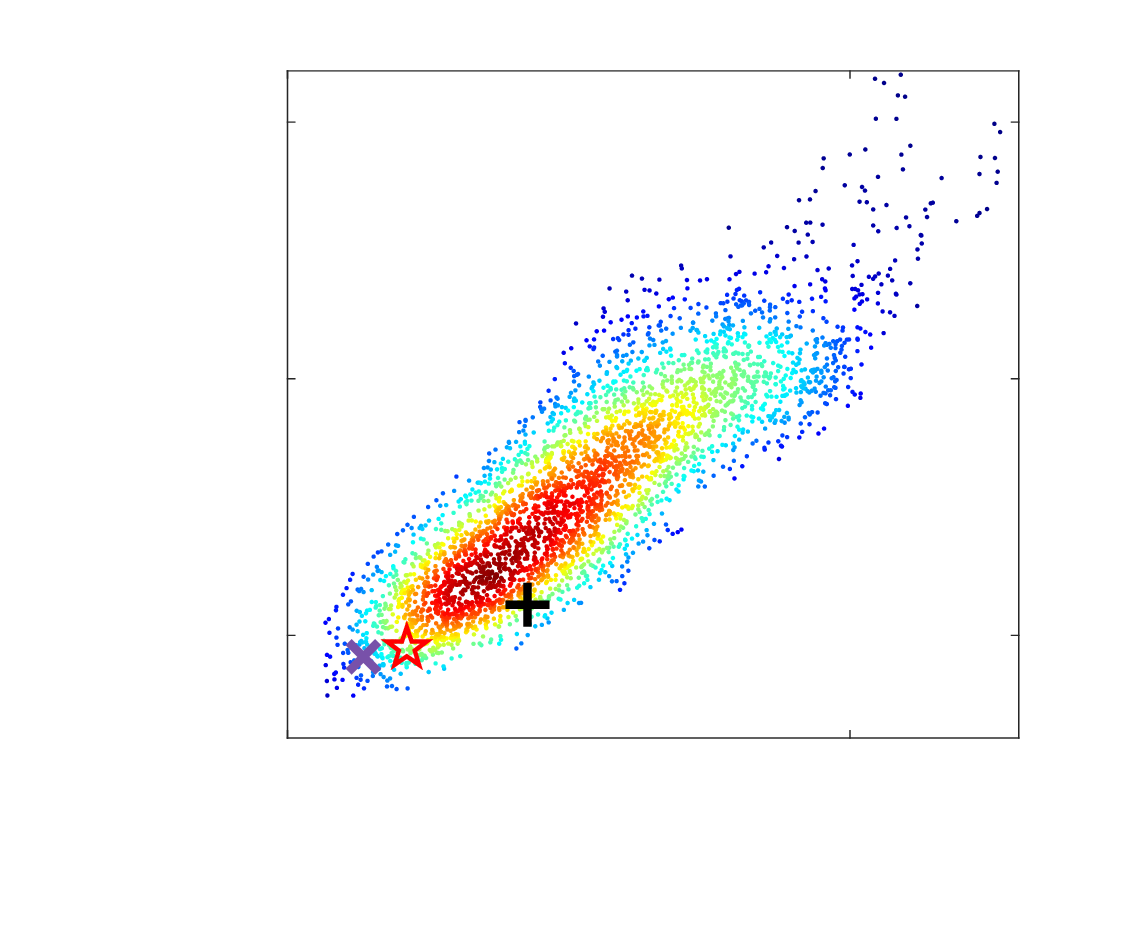}
\end{subfigure}
\begin{subfigure}
    \centering
    \includegraphics[width=0.42\linewidth,trim=0mm 0mm 16mm 0mm,clip]{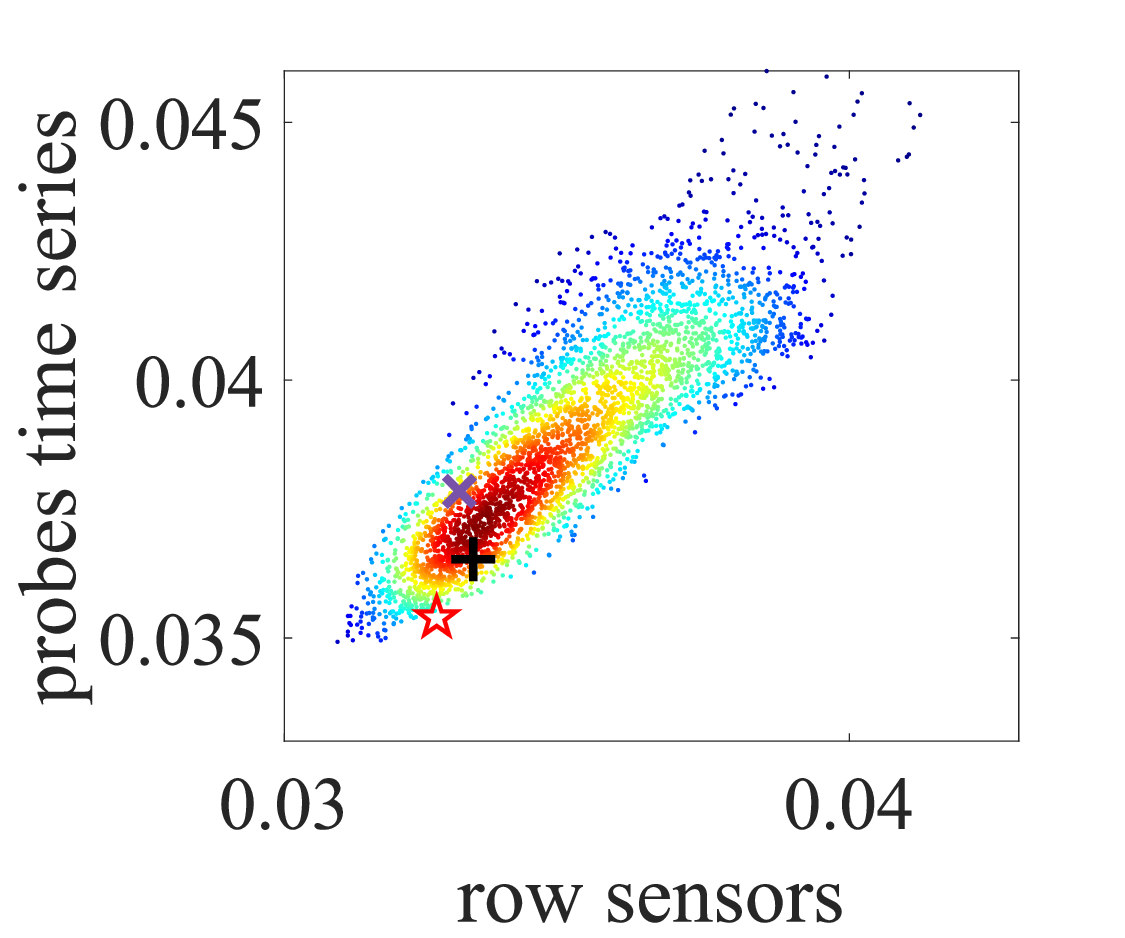}
\end{subfigure}
\begin{subfigure}
    \centering
    \includegraphics[width=0.36\linewidth,trim=24mm 0mm 12mm 0mm,clip]{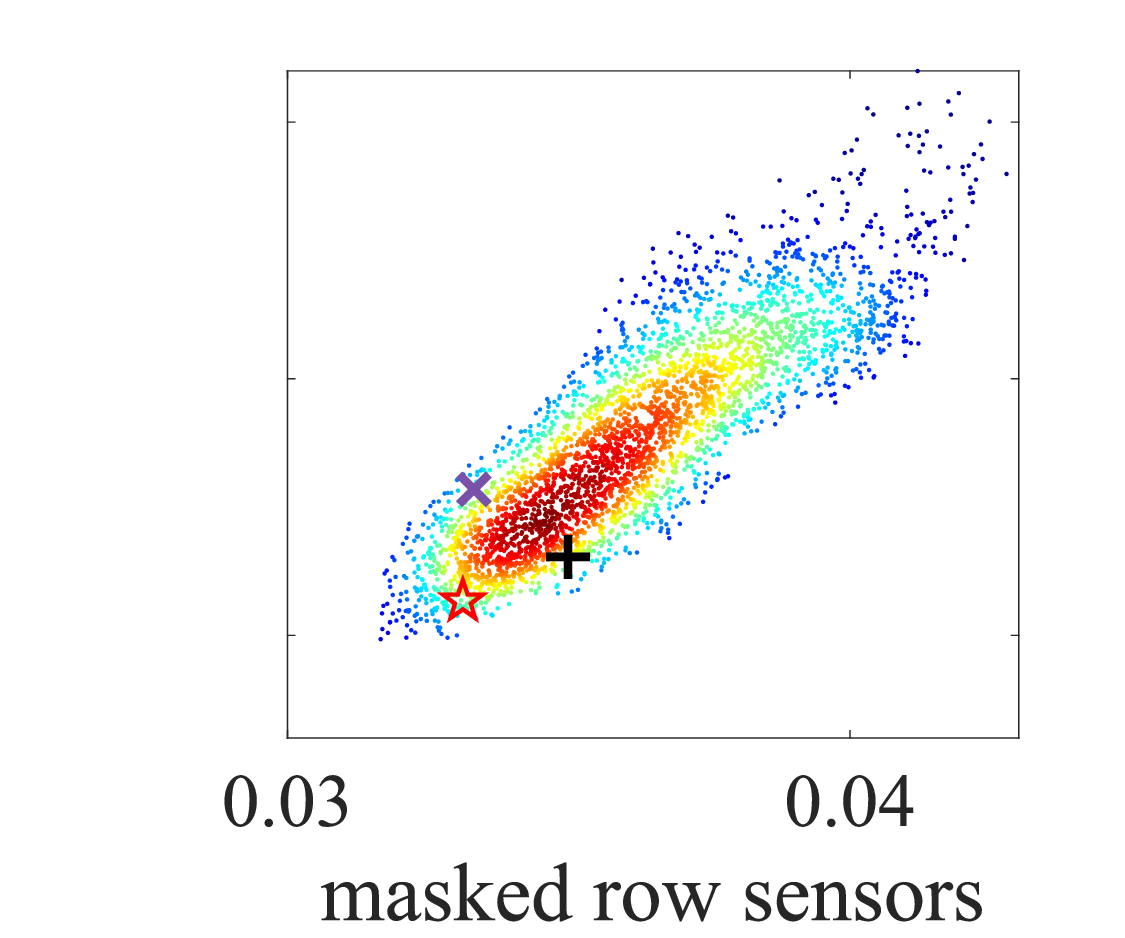}
\end{subfigure}
\caption{The scatter plots of velocity reconstruction error using probes with time-series versus using row sensors (left) and row sensors masked (right). The computation domains are, from top to bottom: 2D-\ac{2C} field in the central plane (top); 2D-3C field in the central plane (central); thin volumetric field (bottom). The colour of the spots represents local density ranging from low (blue) to high (red). The $+$, $\times$ and $\star$ show the result from the placement of equidistant, block-pivoted and optimal combination tested by row sensors or row sensors masked. The density is defined as $\displaystyle\sum_i\rho\Delta s_i = 1$, where the $\Delta s_i$ is any local area on the scatter plots.}
\label{fig: blades_scatter}
\end{figure}

The scatter plots in Fig. \ref{fig: blades_scatter} exhibit similar distributions of the reconstruction error from probes time series versus row sensors and masked row sensors in all tested configurations. This confirms that the offline optimization process is representative of the (more time consuming) online optimization process with probes spanning the volume. It must also be remarked that both types of row sensors outperform either equidistant positioning and block-pivoted QR.

\begin{table*}[]
\centering
\begin{tabu}{|[2pt]l|[2pt]c c c c|[2pt]c|[2pt]}
    \tabucline[2pt]{-}
    \multirow{2}{*}{\diagbox{field type}{probe type}} & \bf{equidistant} & \bf{block-} & \bf{surrogate} & \bf{surrogate masked} & \bf{probes} \\ 
    & \bf{probes} & \bf{pivoted QR} & \bf{row sensors} & \bf{row sensors} & \bf{time series} \\
    \tabucline[2pt]{-}
    \bf{planar PIV}  & 0.0365 & 0.0375 & 0.0360 & 0.0345 & 0.0343 \\
    \hline
    \bf{stereoscopic PIV} & 0.0356 & 0.0346 & 0.0345 & 0.0348 & 0.0338 \\
    \hline
    \bf{thin tomographic PIV:} &  &  &  &  &  \\
    with 3D positioning & 0.0365 & 0.0378 & 0.0353 & 0.0356 & 0.0349 \\
    with 2D positioning & & & 0.0363 & 0.0357 & \\
    \tabucline[2pt]{-}
\end{tabu}

\caption{The error of velocity reconstruction using time-delay embedding and sensors located according to different methods for the experimental dataset of the propeller wake. For the volumetric test case, we consider either exploring the entire 2D donwstream end of the domain (in the table as ``3D positioning'') or solely the central slice (``2D positioning''). The error is defined as in Eq. \ref{eq: error}. In the last column, the optimal case of online optimization with probes time series is included for reference.}
\label{tab: blade_error}
\end{table*}

\section*{Conclusions}
\label{sec:conclusions}
In this paper, we introduced an offline optimization of the probe positioning for flow field estimation targeted at temporal resolution enhancement of \ac{PIV}. The offline position optimization requires solely one non-time-resolved \ac{PIV} sequence measurement. Under the condition that the flow is dominated by advection, time-delay embedding in the probe data can be used to enrich the dataset and improve the accuracy of the reconstruction. Our offline optimization strategy consists of using directly lines of velocity vectors in the main direction of advection from a preliminary snapshot \ac{PIV} experiment, without placing any real probes, to estimate the optimal sensor placement for reconstruction with time-delay embedding. The optimal combination of probe positions is identified through a brute-force combinatorial search. The best combination is then used to locate the probes and perform the final flow estimation experiment with synchronized probe and \ac{PIV} measurements.

We show that row surrogate sensors are in general capable of identifying the most promising locations for estimation of time-resolved fields once real probes are placed. The process is further enhanced by excluding regions with low correlation to potential probe positions, typically found at the downstream edge of the domain in advective flows.

Remarkably, we observed that a simple planar \ac{PIV} experiment is sufficient to locate the sensor position even if the final target is time-resolution enhancement of volumetric measurements in relatively thin domains. 

The offline search of the best probe positions is done by brute-force search. We have explored a modified version of a greedy optimization algorithm based on QR-pivoting to simplify this step further. Nonetheless, we observed that this method is not capable of identifying combinations of probe positions with acceptable accuracy, and it is outperformed by equidistant probe placement in all cases. We envision that further research in this direction is needed. Indeed, brute-force search is made feasible here thanks to the low computational cost of \ac{EPOD}. More complex estimators might require reducing the number of estimations to perform in the offline optimization step. Solutions based on Bayesian optimization, genetic algorithms and reinforcement learning, among others, might be explored in the future. In addition, the flow estimation method might be applied to conditional measurements.

\section*{Acknowledgements}
This project has received funding from the European Research Council (ERC) under the European Union’s Horizon 2020 research and innovation programme (grant agreement No 949085, NEXTFLOW ERC StG). Views and opinions expressed are however those of the authors only and do not necessarily reflect those of the European Union or the European Research Council. Neither the European Union nor the granting authority can be held responsible for them.

\section*{Data availability}
All codes developed in this work are openly available in GitHub, accessible through the link \url{https://github.com/erc-nextflow/sensor_placement_V1}. The dataset are openly available in Zenodo, accessible at \url{https://doi.org/10.5281/zenodo.15114116}

\appendix

\bibliographystyle{elsarticle-harv}






\end{document}